\documentclass[a4paper,aps,prd,10pt,preprintnumbers,showpacs,twocolumn,superscriptaddress,nofootinbib,amsmath,amssymb,floatfix]{revtex4-1}\def\JCAPstyle#1{}
\usepackage{enumitem}
\usepackage{graphicx}
\usepackage[utf8]{inputenc}
\usepackage[T1]{fontenc}
\usepackage{cmap}
\usepackage{subfigure}
\usepackage{color}

\DeclareMathAlphabet{\pazocal}{OMS}{zplm}{m}{n}

\begin{document}

\preprint{APS/-QNM}

\title{Thermodynamic topology of  dyonic AdS black holes with quasitopological electromagnetism in Einstein-Gauss-Bonnet gravity }
\author{Hao Chen}
\email{haochen1249@yeah.net }
\affiliation{School of Physics and Electronic Science, Zunyi Normal University, Zunyi 563006,PR  China}
\author{Meng-Yao Zhang}
\affiliation{College of Computer and Information Engineering,
Guizhou University of Commerce, Guiyang, 550014, China}
\author{Hassan  Hassanabadi}
\affiliation{Department   of   Physics, Faculty of Science,   University   of   Hradec   Kr\'{a}lov\'{e},  Rokitansk\'{e}ho 62, 500   03   Hradec   Kr\'{a}lov\'{e},   Czechia}
\author{Bekir Can L\"{u}tf\"{u}o\u{g}lu}
\affiliation{Department   of   Physics, Faculty of Science,   University   of   Hradec   Kr\'{a}lov\'{e},  Rokitansk\'{e}ho 62, 500   03   Hradec   Kr\'{a}lov\'{e},   Czechia}

\author{Zheng-Wen Long}
\affiliation{College of Physics, Guizhou University, Guiyang, 550025, China}

\begin{abstract}
		\textbf{Abstract:} In this study, we investigate the thermodynamic topology of the high-dimensional dyonic AdS black holes with quasitopological electromagnetism in the Einstein-Gauss-Bonnet background. We first examine the topological charge connected to the critical point and find that the two conventional critical points $CP_{1},CP_{2}$ of the black hole are physical critical point, and the novel critical point $CP_{3}$ that lacks the capability to minimize the Gibbs free energy ($\alpha=0.5$). The critical points $CP_{1}$ and $CP_{2}$ are observed to occur at the maximum extreme points of temperature in the isobaric curve, while the critical point $CP_{3}$, emerges at the minimum extreme points of temperature. Furthermore, the number of phases at the novel critical point exhibits an upward trend, followed by a subsequent decline at the conventional critical points. With the increase of the coupling constant ($\alpha = 1$), although the system has three critical points, only the conventional $CP_{1}$ is a (physical) critical point, and the conventional $CP_{2}$ serves as the phase annihilation point.  This means that the coupling constant $\alpha$ has significant impact on the phase structure. Additionally, we regard dyonic AdS black holes as a topological defect within the thermodynamic space, our findings indicate that alterations in pressure can result in the system exhibiting distinct points of generation and annihilation. However, the total topological number of black holes in different dimensions is $1$, the system shares a similar topological classification as the charged RN-AdS black holes. The discovery we have made provides a crucial component in understanding the thermodynamic topology of dyonic AdS black holes.
\end{abstract}
	
	\maketitle
	
\section{Introduction}
The official announcement on February 11, 2016 of the detection of gravitational waves by the LIGO represents a groundbreaking achievement that provides significant validation for Einstein's general relativity. This momentous discovery not only confirms the existence of black holes but also verifies the presence of gravitational waves \cite{ch1,ch2,ch3,ch4}. The emergence of black hole (BH) thermodynamics stems from the application of general laws in thermodynamics to the general relativity, with the objective of investigating the characteristics inherent to black holes. The fundamental concept involves regarding black holes as a system governed by thermodynamic principles, wherein mass corresponds to energy, entropy is directly proportional to the surface area, and temperature is represented by surface gravity. In 1974, Hawking made a groundbreaking discovery that black holes exhibit thermal radiation \cite{ch5}, this revelation not only resolves the inherent contradictions within the thermodynamics of black holes but also successfully establishes a profound connection between gravitational theory, quantum mechanics, and thermodynamics. The phenomenon of phase transition is commonly observed in thermodynamic systems, wherein the Schwarzschild black hole, when considered as a thermodynamic system within the framework of AdS, exhibits an unstable state from a thermodynamic perspective at low temperatures. However, with increasing temperature, the free energy is reduced, and finally the black hole reaches stability.

Kastor and Ray proposed the concept of regarding the cosmological constant $\Lambda$ as a variable thermodynamic pressure, and suggested that the volume of the black hole in AdS space can be considered as its conjugate quantity, while the black hole's mass is interpreted as the spacetime's enthalpy, rather than its internal energy \cite{ch8}. This notion was further expanded to Lovelock theory and applied to black holes, which demonstrates that Lovelock coupling serves as an extension of the first law of thermodynamics \cite{ch9}. Additionally, both the free energy  and of Arnowitt-Deser-Misner (ADM) mass AdS black holes were derived \cite{ch10}. Furthermore, investigations on dilaton-AdS black holes have revealed that their mass represents the system's enthalpy \cite{ch11}, the phenomenon has led to a surge in research on thermodynamic phase transitions in AdS black holes. For instance, the charged RN-AdS black hole in four-dimensional spacetime was examined by  Kubiznak and Mann, who identified a significant transition between small and large black holes, which bears resemblance to the gas-liquid phase transition observed in van der Waals fluids \cite{ch17}. The phase transition in nonlinear electromagnetic field was investigated by Hendi and Vahidinia through considering the Einstein gravitational solution under the background of generalized Maxwell theory \cite{ch18}. Critical phenomena and critical exponents near the critical point of Gauss-Bonnet black hole were studied by Wei et al, who also examined the phenomenon of phase transition in spin Kerr black holes with a high number of dimensions \cite{ch19,ch20}. Zou et al. explored P-V critical behavior of Born-Infeld Anti-deSitter black hole and Gauss-Bonnet AdS black hole \cite{ch21,ch22}. It is well known that the typical first-order phase transition has only one critical point. However, the existence of multiple critical points is observed in certain categories of black holes, such as, the occurrence of reentrant phase transition \cite{ch24,ch25}. For the intriguing solution of black hole with quasi-topological electromagnetism (QTE) \cite{sci1}, this theory presents a novel higher-order expansion formulated using the bilinear norm of Maxwell's theory.  In certain specific circumstances, the presence of extra terms does not have any impact on the energy-momentum tensor and the Maxwell equation. The presence of triple points in dynamic AdS black holes with QTE \cite{ch26,ch27}, and the $\lambda$ phase transition observed in Horava gravity \cite{ch28}, among others. A complementary aspect to the investigation of black hole criticality is the concept of thermodynamic topology. The Duan's $\varphi$-mapping theory proposed is utilized to explore the criticality of black holes, resulting in distinct characteristics of topological charge exhibited by critical points within the thermodynamic space, these critical points can be classified into conventional  and novel types \cite{cq4}. Furthermore, Wei et al. employed the topological method to investigate thermodynamic properties of black holes, this approach enables analysis of stability by considering the winding number associated with defect \cite{we1}. Recently, Fang et al. proposed that the winding numbers can be determined by calculating the residue of the isolated first-order pole of the characterized functions constructed using off-shell free energy. This approach effectively enables the classification of black hole topology \cite{l11}.

The theory of extended gravity encompasses Einstein-Gauss-Bonnet (EGB) gravity, which was first proposed by Cisterna et al \cite{rna1}, by using this model for other constructions, specifically, extended black strings and p-branes and diverse compactifications with Lovelock fluxes \cite{rna2,rna3}. EGB gravity provides novel perspectives on the characteristics of black holes \cite{ch29,ch30,ch31,ch32,ch33,ch34,ch35}. Through an examination of the thermodynamic characteristics of the Born-Infeld AdS black hole within the framework of 4D EGB gravity, it has been observed that there exist three critical points within the black hole, these critical points exhibit numerous intriguing properties when subjected to EGB gravity \cite{ch25}. Recently, the QTE is characterized by the square norm of the topological 4-form $F \wedge F$, indicating that the system's energy-momentum tensor corresponds to a perfect fluid with isotropic properties \cite{sci1}. Based on this, the thermodynamic phase transition of dyonic AdS black holes with QTE is discussed in detail. By changing the coupling parameters, a triple points phase structure is observed \cite{ch27}. Additionally, through an examination of the interplay between QTE and EGB gravity, Sekhmani et al. derived a solution for dyonic AdS black holes with a high number of dimensions. Furthermore, they conducted a comprehensive analysis on how the coupling constant $\alpha$ influences both the energy emission rates and black hole shadow \cite{ch37}. On this basis, a comprehensive analysis is conducted on the phase transition of the dyonic AdS black holes with QTE in the EGB background, and it is found that the system parameters will appear triple points in special regions \cite{cq3}. In this work, we will use the topological method to investigate the critical points, and further understand the influence of GB coupling constant on the phase structure of dyonic AdS black holes. The structure of the manuscript is as follows: In section \ref{sec2}, we give a concise overview of the thermodynamic properties of dyonic AdS black holes with QTE in the EGB background and present a comprehensive explanation of Duan's $\varphi$-mapping theory and  the off-shell method. In section \ref{sec3}, we will use the topological method to classify the critical points. In section \ref{sec4}, we determine the characteristics of the critical point by analyzing the isobaric curve. In section \ref{sec5}, we examine dyonic AdS black holes as a defect in the thermodynamic space, aiming to comprehend the local and global topological structure of this defect. Section \ref{sec6} contains the remarks and conclusions.
\section{Dyonic AdS black holes and topology}\label{sec2}
We first consider the d-dimensional action involving EGB gravity and QTE which can be formulated in a minimal coupling manner \cite{rna1,ch37}
\begin{equation}
S_D=\frac{1}{16 \pi} \int d^d x \sqrt{-g}\left(\mathcal{R}-2 \Lambda+\alpha \mathcal{G}+\mathcal{L}_{\mathrm{QE}}\right),
\end{equation}
wherein, $\Lambda=-\frac{(d-1)(d-2)}{2l^{2}}$ represents the cosmological constant within the AdS space, $\alpha$ is the symbol used to denote the GB coupling constant, which possesses dimensions of [length]$^{2}$. Here, the GB term can be expressed as
\begin{equation}
\mathcal{G}=\mathcal{R}^{2}-4\mathcal{R}^{\mu\nu}\mathcal{R}_{\mu\nu}+\mathcal{R}^{\mu\nu\rho\sigma}\mathcal{R}_{\mu\nu\rho\sigma}. \end{equation}
The matter field, a fundamental concept in the realm of physics, can be characterized by an electromagnetic Lagrangian with quasi-topological properties and represented as
\begin{align}
\mathcal{L}_{\rm{QE}}=-(\frac{1}{4}\mathcal{F}^{2}+\frac{1}{2p!}\mathcal{H}^{2}+\beta\mathcal{L}_{\rm{int}}),
\end{align}
where $p=d-2$, $\mathcal{F}^{2}=\mathcal{F}_{\mu\nu}\mathcal{F}^{\mu\nu}$ and $\mathcal{H}^{2}=\mathcal{H}_{\rho_{1}\ldots\rho _{p}}\mathcal{H}^{\rho_{1}\ldots\rho _{p}}$ , in which, according to Ref. \cite{rna1}, the gauges fields $\mathcal{F}_{\mu\nu}$ and $\mathcal{H}_{\rho_1...\rho_p}$ are considered purely electric and magnetic, respectively. Here,  $\beta$ means the coupling constant associated with the dimension [length]$^{2}$.
In this case the only non-vanishing terms are $\mathcal{F}_{\mu\nu}\mathcal{F}^{\mu\nu}$, $|\mathcal{H}_{\left[p\right]}|^2$, and $|\mathcal{F}\mathcal{H}_{\left[D\right]}|^2$.
Additionally, the interaction term reads
\begin{align}
	\mathcal{L}_{\rm{int}}=\delta_{\gamma_{1}\ldots\gamma _{D}}^{\lambda_{1}\ldots\lambda _{D}}\mathcal{F}_{\lambda_{1}\lambda_{2}}\mathcal{H}_{\lambda_{3}\ldots\lambda _{D}}\mathcal{F}^{\gamma_{1}\gamma_{2}}\mathcal{H}^{\gamma_{3}\ldots\gamma _{D}},
\end{align}
the above action, referring to a specific mathematical calculation or physical process, leads to the emergence of a BH solution, which can be described by a metric
\begin{align}
	ds^2=-f(r)dt^2+f(r)^{-1}dr^2 +r^2 d\Omega_{d-2} ^2 , \
\end{align}
where $d\Omega_{d-2}^{2}$ represents the unit sphere in $(d-2$) dimensions, the metric function is provided by
\begin{equation}
f(r)=1+\frac{r^2}{2 \tilde{\alpha}}\left(1-\sqrt{1+\frac{4 \tilde{\alpha} m}{r^{d-1}}+\frac{8 \tilde{\alpha} \Lambda}{(d-2)(d-1)}-\varrho}\right),
\end{equation}
with
\begin{equation}
\varrho=\frac{2 \tilde{\alpha}}{d-3}\left(Q_m^2+Q_e^2 \xi\right),
\end{equation}
 here $\alpha$ is the GB coupling constant, the hypergeometric function, denoted by $\xi$, can be expressed as
\begin{equation}
\xi=\ _2F_1\left[1,\frac{d-3}{2(d-2)};\frac{7-3d}{4-2d};\frac{-8\beta Q_{m}^{2}\Gamma(d-1)^{2}}{r^{2d-4}}\right],
\end{equation}
\begin{equation}
\widetilde{\alpha}=(d-4)(d-3)\alpha,
\end{equation}
wherein, $Q_{e}$ and $Q_{m}$ are essential parameters that indicate the electrical and magnetic charges of  black holes, respectively. These parameters are critical in comprehending the characteristics of black holes. $m$ denotes the mass of the solution within a designated parameter domain, serving as the constant of integration derived from the boundary conditions \cite{aj1}. In this case, ADM mass is given by
\begin{align}
	M=\frac{(d-2)\mathcal{V}_{d-2}}{16\pi}m,
\end{align}
with
\begin{equation}
\mathcal{V}_{d-2}=\frac{2\pi^{\frac{d-1}{2}}}{\Gamma\left( \frac{d-1}{2}\right) }.
\end{equation}
The equation $f\left(r_+\right)=0$ determines the value of $\mathrm{BH}$ horizon radius $r_+$, which can be obtained through solving this equation. The concept of thermodynamic pressure is closely related to the cosmological constant $\Lambda$, specifically, $P=-\frac{\Lambda}{8\pi}$ \cite{ch8}. Hence, the mass of $\mathrm{BH}$ can be represented in relation to the horizon radius $r_+$ as
\begin{equation}\label{wei6}
\begin{aligned}
& M=\frac{1}{32\left(3-4 d+d^2\right) \pi} r_{+}^{-5-d} \mathcal{V}_{d-2} \\
& \quad \times\left[(d-1) Q_m^2 r_{+}^8+2(d-3) r_{+}^{2 d}\right. \\
& \quad \times\left(\left(2-3 d+d^2\right) r_{+}^2\right. \\
& \left.\quad+\left(24-50 d+35 d^2-10 d^3+d^4\right) \alpha+16\pi r_{+}^4 P\right) \\
& \quad+(d-1) Q_e^2 r_{+}^8{ }_2 F_1 \\
& \left.\quad \times\left(1, \frac{d-3}{2(d-2)}, \frac{7-3 d}{4-24},-8 \beta Q_m^2 r_{+}^{4-2 d} \Gamma(d-1)^2\right)\right].
\end{aligned}
\end{equation}
Furthermore, by utilizing the definition of Hawking temperature as $T=\frac{f^{\prime}\left(r_+\right)}{4 \pi}$, we can obtain
\begin{equation}
T_{+}=\frac{t_1}{t_2},
\end{equation}
where
\begin{equation}
\begin{aligned}
t_1= & r_{+}^{-2 d-1}\left(8 \beta Q _ { m } ^ { 2 } r _ { + } ^ { 4 } \Gamma ( d - 1 ) ^ { 2 } \left(-2 r_{+}^{2 d}\left(2 \Lambda r_{+}^4\right.\right.\right. \\
& \left.\left.-(d-3)(d-2)\left(\alpha(d-5)(d-4)+r_{+}^2\right)\right)-q_m^2 r_{+}^8\right) \\
& -r_{+}^{2 d+8}\left(Q_e^2+Q_m^2\right)-2 r_{+}^{4 d}\left(2 \Lambda r_{+}^4\right. \\
& \left.\left.-(d-3)(d-2)\left(\alpha(d-5)(d-4)+r_{+}^2\right)\right)\right),
\end{aligned}
\end{equation}
\begin{equation}
\begin{aligned}
t_2= & 8 \pi(d-2)\left(2 \alpha(d-4)(d-3)+r_{+}^2\right) \\
& \times\left(8 \beta Q_m^2 r_{+}^4 \Gamma(d-1)^2+r_{+}^{2 d}\right).
\end{aligned}
\end{equation}
The Bekenstein-Hawking entropy in \cite{cq3} can be expressed as 
\begin{equation}
S=\frac{\pi^{\frac{d-1}{2}} r_+^{d-2}\left[1+2(d-3)(d-2) \alpha r_+^{-2}\right]}{2 \Gamma\left(\frac{d-1}{2}\right)},
\end{equation}
other thermodynamic quantities, including the volume $V$, electric potential $\Phi_e$, and magnetic potential $\Phi_m$, can be performed by
\begin{equation}
V=\frac{2 \pi^{\frac{d-1}{2}} r_+^{d-1}}{(d-1) \Gamma\left(\frac{d-1}{2}\right)},
\end{equation}
\begin{equation}
\Phi_e=\left(\frac{\partial M}{\partial Q_e}\right)=\frac{\pi^{\frac{d-3}{2}} r_+^{3-d} Q_e \xi}{8(d-3) \Gamma\left(\frac{d-1}{2}\right)},
\end{equation}
\begin{equation}
\begin{aligned}
\Phi_m&=\left(\frac{\partial M}{\partial Q_m}\right)\\
&=\frac{\pi^{\frac{d-3}{2}} r_+^{3-d}\left[2(d-2) Q_m^2+(d-3) Q_e^2\left(r_+^{2 d} \mathcal{\varsigma}^{-1}-\xi\right)\right]}{16(d-3)(d-2) Q_m \Gamma\left(\frac{d-1}{2}\right)},
\end{aligned}
\end{equation}
where
\begin{equation}
\mathcal{\varsigma}=r_+^{2d}+8 r_+^4 \beta Q_m^2 \Gamma(d-1)^2.
\end{equation}
Similarly, the conjugate quantity to the coupling constant $\alpha$ and $\beta$ can be formulated as
\begin{equation}
\begin{aligned}
\Phi_\alpha& \equiv\left(\frac{\partial M}{\partial \alpha}\right)\\
&=\frac{\pi^{\frac{d-3}{2}} r_+^{d-5} \mathcal{\varsigma}_{1}}{8 \alpha \Gamma\left(\frac{d-1}{2}\right)}-\frac{(d-3)(d-2) \pi^{\frac{d-1}{2}} r_+^{d-4} T}{\Gamma\left(\frac{d-1}{2}\right)},
\end{aligned}
\end{equation}
where
\begin{equation}
\mathcal{\varsigma}_{1}=(d-4)(d-3)(d-2) \alpha
\end{equation}
\begin{equation}
\Phi_\beta =\left(\frac{\partial M}{\partial \beta}\right)
=\frac{\pi^{\frac{d-3}{2}} r_+^{3-d} Q_e^2\left(r_+^{2 d} \varsigma-\xi\right)}{32(d-2) \beta \Gamma\left(\frac{d-1}{2}\right)}.
\end{equation}
By utilizing this information, the generalized Smarr relation is given by
\begin{equation}
\begin{aligned}
M & =\frac{d-2}{d-3} T S-\frac{2}{d-3} P V+\Phi_e Q_e+\Phi_m Q_m \\
& +\frac{2}{d-3}\left(\alpha \Phi_\alpha+\beta \Phi_\beta\right).
\end{aligned}
\end{equation}
To gain a deeper comprehension of the phase structure, we will employ the thermodynamic topology approach to classify the critical points. Drawing inspiration from \cite{cq4}, which implies the allocation of topological charge to these critical points, we denote the temperature as $T=T\left(S, P, x^i\right)$. Based on this premise, the determination of the critical point of the black hole can be expressed as
\begin{equation}
\left(\partial_S T\right)_{P, x^i}=0, \quad\left(\partial_{S,S} T\right)_{P, x^i}=0.
\end{equation}
The elimination of thermodynamic pressure leads to the derivation of a novel temperature function from equation (18), the next step entails the construction of a Duan's potential \cite{cq4}
\begin{equation}
\Phi=\frac{1}{\sin \theta} T\left(S, x^i\right),
\end{equation}
here, the presence of an additional factor $1/\sin \theta$ contributes to the topological analysis. The application of Duan's $\phi$-mapping theory is employed to introduce a new vector field denoted as $\phi=(\phi^S,\phi^\theta)$, where
\begin{equation}
\phi ^S=(\partial _S \Phi )_{\theta ,x_i}, \phi ^\theta=(\partial _\theta \Phi )_{ S,x_i}.
\end{equation}
Inspired by \cite{cq5}, in the case where $\theta$ equals $\frac{\pi}{2}$, the $\phi$ component of the vector field remains consistently at zero, allowing for a description of the topological flow as follows:
\begin{equation}
J^\mu=\frac{1}{2\pi } \epsilon ^{\mu \nu \lambda }\epsilon _{ab}\partial_{ \nu }n^a\partial_{ \lambda  }n^b,
\end{equation}
where $\partial_{ \nu }=\frac{\partial}{\partial x^{\nu}}$ and $x^{\nu}=(t, r, \theta)$. The unit vector $n$ is  represented  by $n=(n^1, n^2)$, where $n^1=\frac{\phi ^S}{\left \| \phi  \right \| }$, $n^2=\frac{\phi ^\theta }{\left \| \phi  \right \| } $. The fulfillment of $\partial_\mu J^\mu=0$ is a prerequisite for the topological current. In addition, the topological charge linked to a parameter region $\sum $ can be expressed as follows:
\begin{equation}
Q_t=\int_{\sum} j^0 d^2 x=\sum_{i=1}^N w_i,
\end{equation}
the winding number $w_i$ determines the topological charge $Q$ at zero points $(\phi ^a(x^i)=0)$, with negative or positive values corresponding to a winding number of $-1$ or $+1$ respectively. The summation of charges at all critical points within a thermodynamic system defines the total topological charge $Q_t$, this fundamental concept holds great significance in the analysis of thermodynamic systems as it aids in comprehending their global characteristics and facilitating classification. The exploration of the topology of black holes has been broadened to include various classifications of black holes \cite{o1,o2,o3,o4,o5,o6,o7,o8,o9,o10,o12}.

A recent study proposes an alternative approach to incorporate topology into the black hole thermodynamics \cite{we1}, this perspective implies that defects can be interpreted as black hole solutions. The total winding number, referred to as the topological number, is employed in the classification of distinct black hole solutions. To begin with, we introduce a generalized free energy denoted as $F$, which is represented by
\begin{equation}\label{wei8}
\mathcal{F}=E-\frac{S}{\tau}.
\end{equation}
In this context, $E$ and $S$ represent energy and entropy respectively, while $\tau$ is a dimensionless quantity indicating time. The vector field $\phi$ is obtained from
\begin{equation}\label{wei10}
\phi=\left(\frac{\partial \mathcal{F}}{\partial r_{+}},-\cot \Theta \csc \Theta\right).
\end{equation}
The vector $\phi$ is centered at $\Theta=\pi / 2$, and its unit vector properties are defined by
\begin{equation}
n^a=\frac{\phi^a}{\|\phi\|} \quad(a=1,2) \quad \text { and } \quad \phi^1=\phi^{r_{+}}, \quad \phi^2=\phi^{\Theta}.
\end{equation}
The identification of the zero points  of $n^1$ relies on the specified value of $\tau$, and the calculation of winding numbers is conducted for zero point, the total  winding  number can be obtained by adding up the individual winding numbers of each branch of the black hole. Up to now, the investigation of topological defects has been expanded to encompass diverse types of black holes \cite{sa1,ad1,ad2,l1,l2,l3,l4,l5,l7,l9,l10,l12,l13,l14,l15,l16,l16.1,l17,l18,l19,l20,l21,l22,l23,l24,meng1,meng2}.
\begin{figure}
		\centering
		\includegraphics[scale = 0.25]{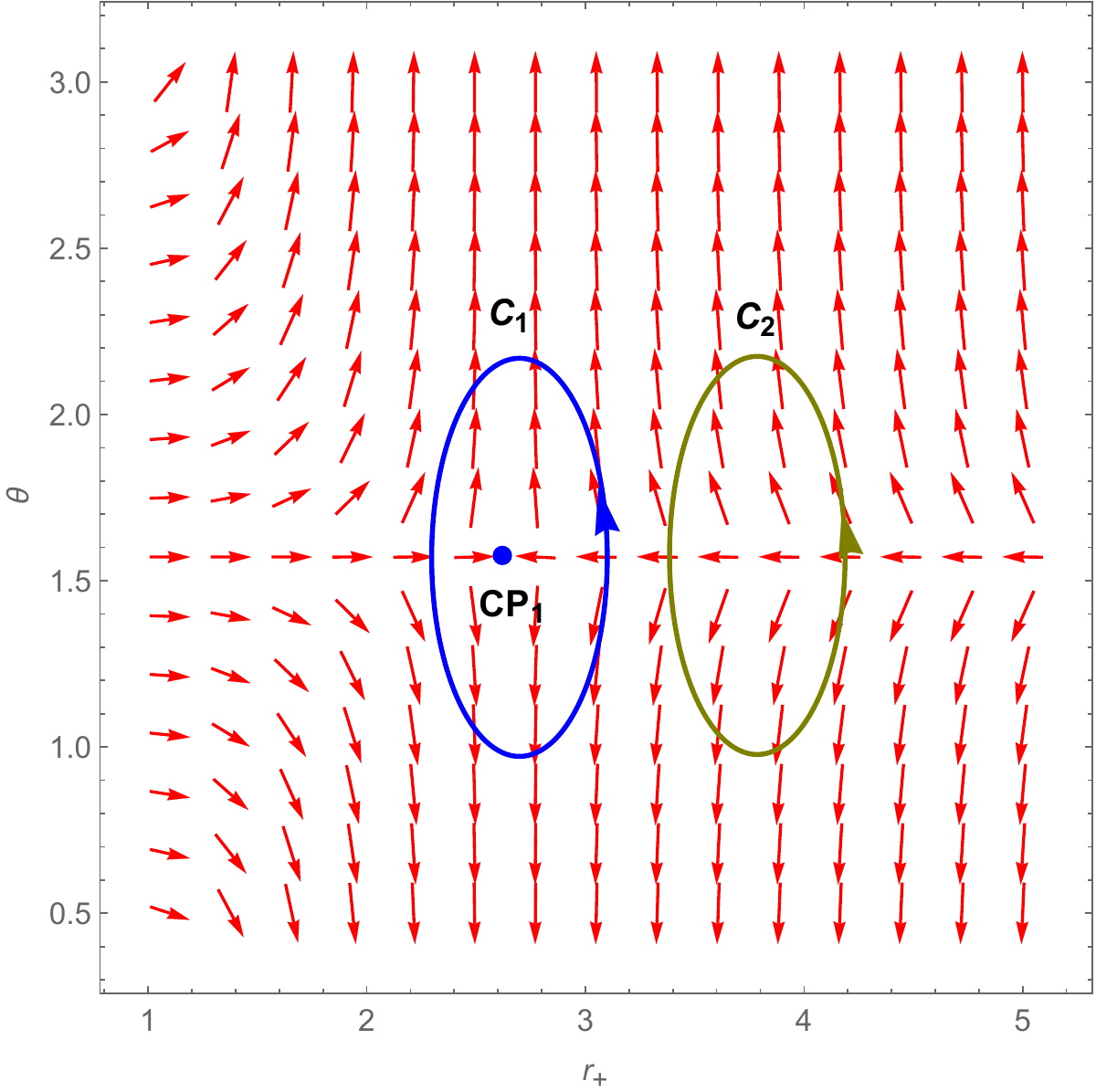} \hspace{-0.5cm}
		\includegraphics[scale = 0.28]{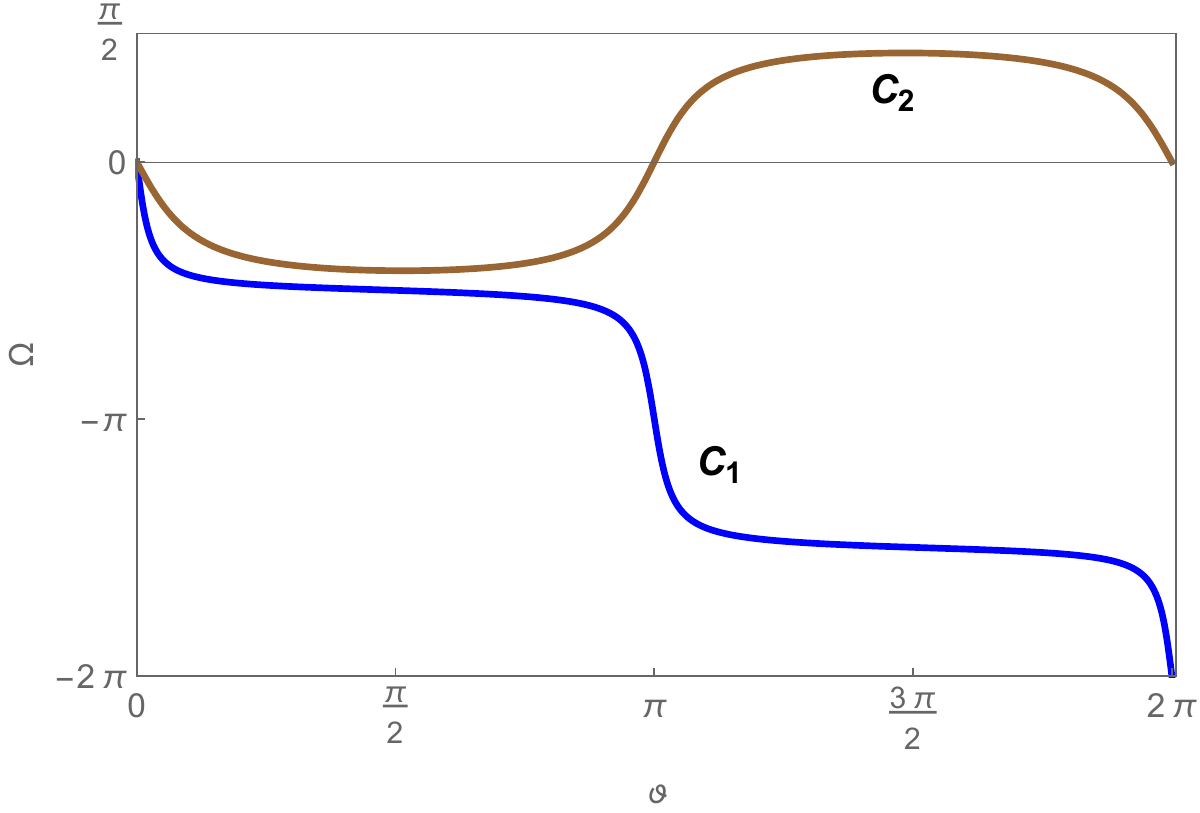} \vspace{-0.5cm}
        \caption{$d=5$: \textbf{ Up panel:} The red arrows represent the vector field $n$ for dyonic AdS black holes with QTE in the EGB background. The blue dot are $(r, \theta) =(2.658528, \pi/2 )$, which represent the critical points $CP_1$. \textbf{Down panel:} The deflection angle $\Omega (\vartheta)$ as a function of $\vartheta$ for contours $C_1$ (blue curve), $C_2$ (brown curve).}
		\label{fig:WKBL1}
	\end{figure}
\begin{figure}
		\centering
		\includegraphics[scale = 0.22]{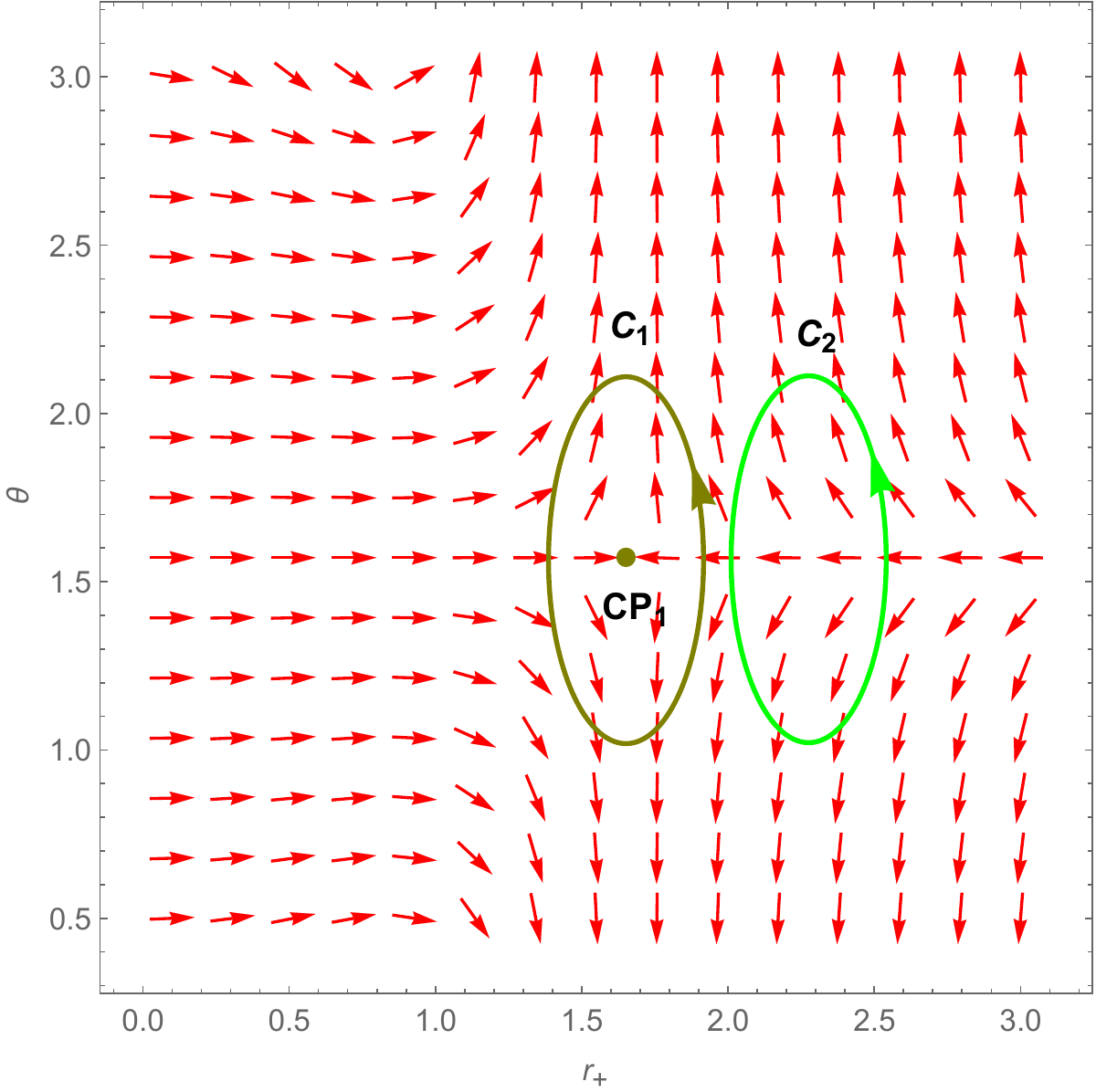} \hspace{-0.5cm}
		\includegraphics[scale = 0.24]{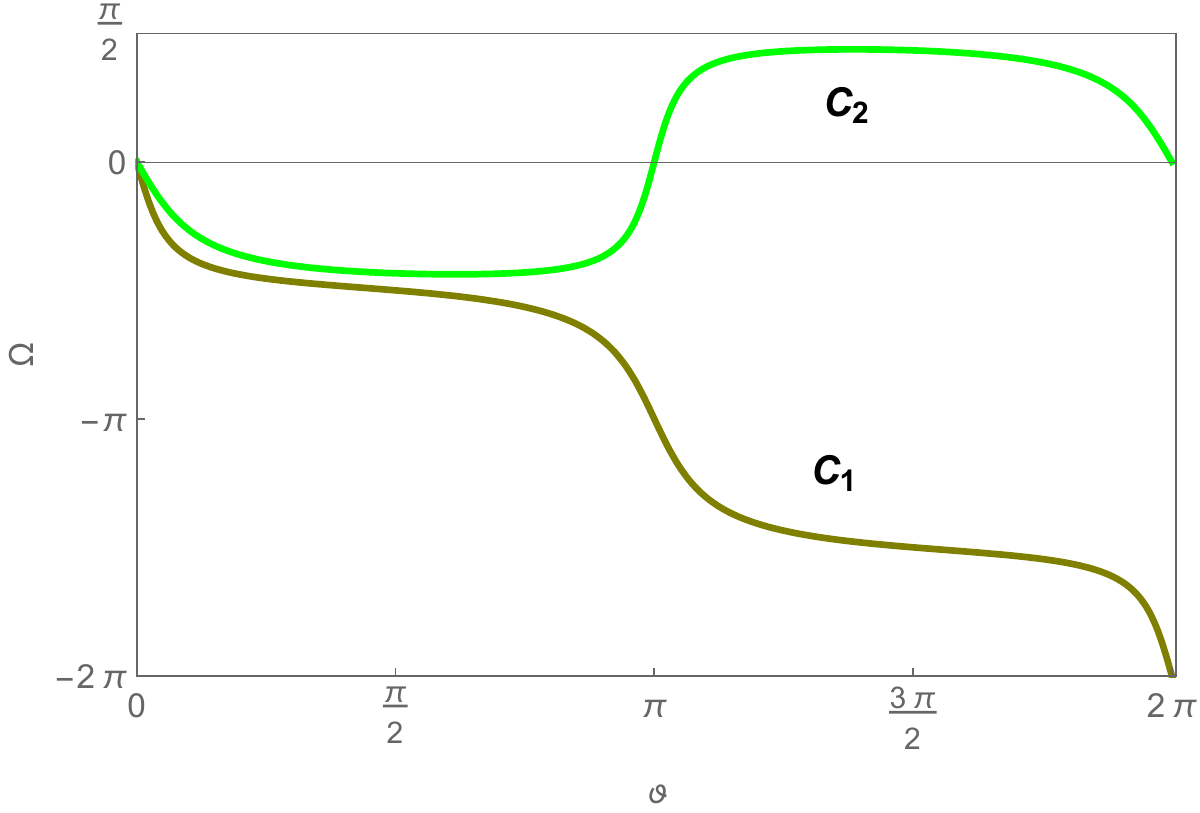} \vspace{-0.5cm}
        \caption{$d=6, \alpha=0.1$: \textbf{ Up panel:} The red arrows represent the vector field $n$ for dyonic AdS black holes with QTE in the EGB background. The green dot are $(r, \theta) =(1.711584, \pi/2 )$, which represent the critical points $CP_1$. \textbf{Down panel:} The deflection angle $\Omega (\vartheta)$ as a function of $\vartheta$ for contours $C_1$ (green curve), $C_2$ (brown curve).}
		\label{fig:WKBL2}
	\end{figure}
\section{Topology of critical points}\label{sec3}
In this section, we will investigate the  categorization of critical points for black holes.  By analyzing the metric of black holes in equation (26), we can deduce the thermodynamic function representing temperature without pressure
\begin{figure}
		\centering
		\includegraphics[scale = 0.26]{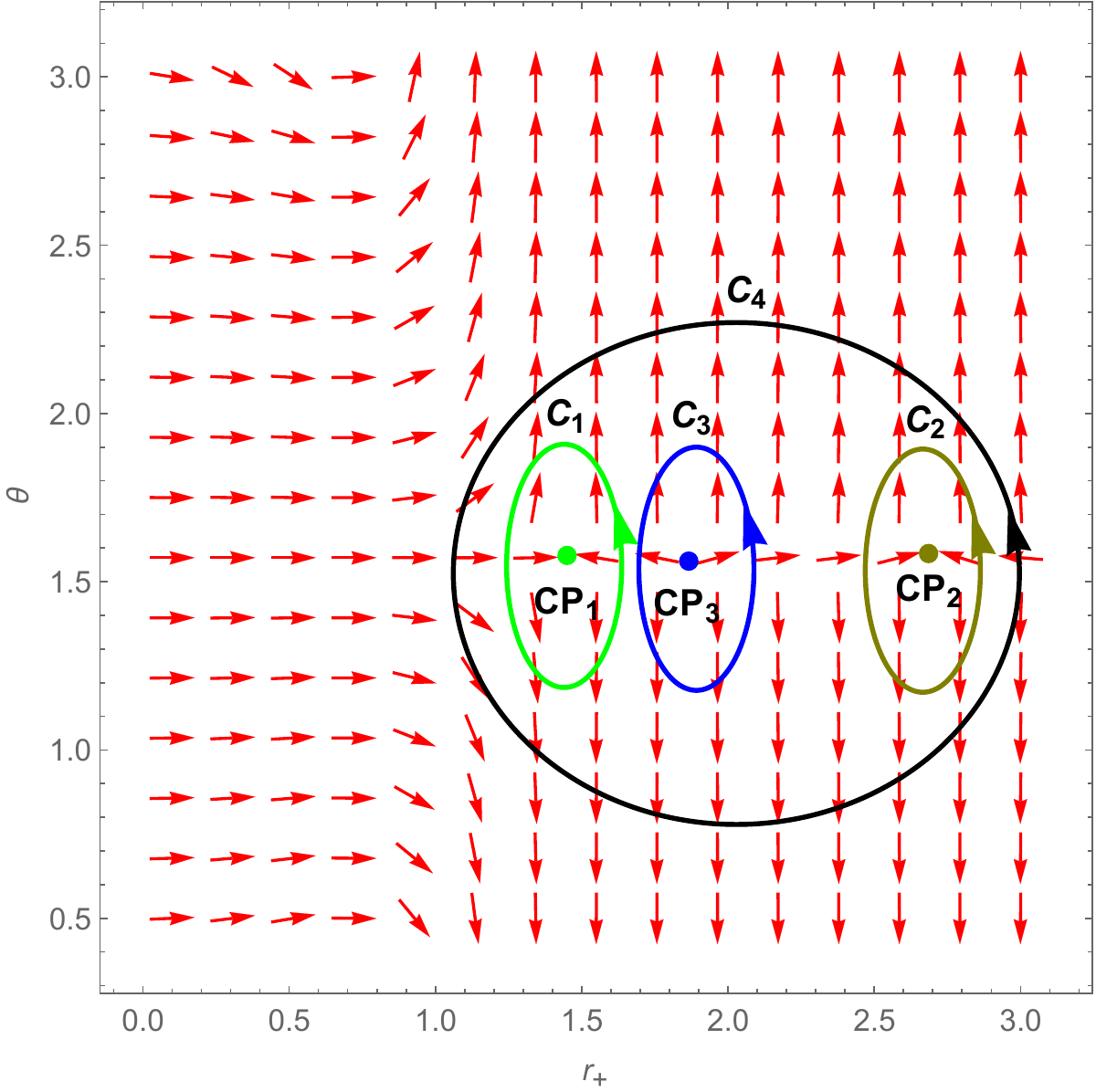} \hspace{-0.5cm}
		\includegraphics[scale = 0.32]{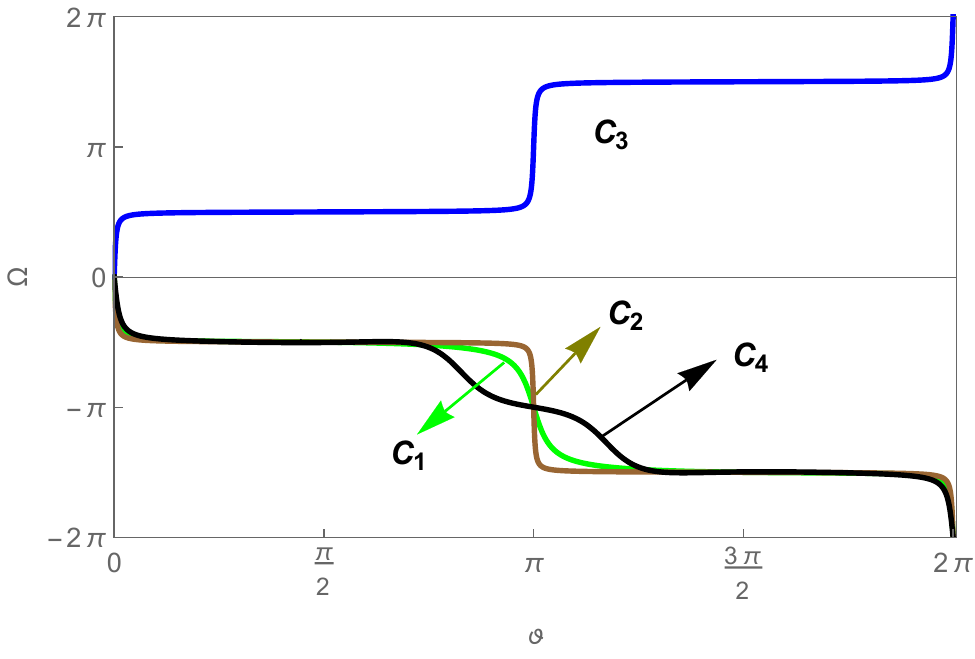} \vspace{-0.5cm}
        \caption{$(d=6,\alpha=0.5,\beta=0.1,Q_{e}=1,Q_{m}=2)$: \textbf{ Up panel:} The red arrows represent the vector field $n$ for dyonic AdS black holes with QTE in the EGB background. The colored dots are $(r, \theta) =( 1.489048, \pi/2 )$, $(r, \theta) =(2.706456, \pi/2 )$ and $(r, \theta) =(1.873569, \pi/2 )$ which represent the critical points $CP_1$, $CP_2$  and $CP_3$, respectively. \textbf{Down panel:}The deflection angle $\Omega (\vartheta)$ as a function of $\vartheta$ for contours $C_1$ (green curve), $C_2$ (brown curve) and $C_3$ (blue curve) and $C_4$ (black curve).}
		\label{fig:WKBL3}
	\end{figure}
\begin{table}[]
	\centering
	\caption{Critical values at $Q_{e}=1,Q_{m}=2, \beta=0.1 $ for dyonic AdS black holes with QTE in the EGB background.}
\begin{tabular}{|lc|l|l|l|l}
 \hline   $$                                  &      & $CP_1$    & $CP_2$    & $CP_3$ \\
\hline                                             & $r_c$  & 2.658528 & - & -     \\
($\alpha=0.5, d=5$)   & $T_c$  & 0.063461 & - & -   \\
                                                  & $P_c$  & 0.006193 & - & -         \\
\hline                                             & $r_c$  & 1.711584 & - & -            \\
($\alpha=0.1, d=6$)           & $T_c$  & 0.138928& - &  -           \\
                                                   & $P_c$  &  0.027903& - &-            \\\hline                                             & $r_c$  & 1.489048 & 2.706456 & 1.873569            \\
($\alpha=0.5, d=6$)           & $T_c$  & 0.064736& 0.064872 &  0.064642           \\
                                                   & $P_c$  &  0.006536& 0.006576 &0.006353            \\
\hline                                             & $r_c$  & 1.008427 & 3.609964 & 3.284875   \\
($\alpha=1.0, d=6$)   & $T_c$  & 0.055925 & 0.045933  &  0.045928    \\
                                                   & $P_c$  & 0.050611& 0.003311 & 0.003308  \\
 \hline
 & $r_c$  & 0.929668 & - & -            \\
($\alpha=0.5, d=7$)           & $T_c$  & 0.113123& - &  -           \\
                                                   & $P_c$  &  0.288091& - &-           \\
\hline                                             & $r_c$  & 0.864544 &  -      & -            \\
($\alpha=0.5, d=8$)                        & $T_c$  & 0.173732 &  -      & -            \\
                                                   & $P_c$  & 1.386830 &  -      & -         \\
                                                   \hline                                             & $r_c$  & 0.845743 &  -      & -            \\
($\alpha=0.5, d=9$)                        & $T_c$  & 0.235677 &  -      & -            \\
                                                   & $P_c$  & 3.856570 &  -      & -         \\ \hline
\end{tabular}
\end{table}
\begin{figure}
		\centering
		\includegraphics[scale = 0.25]{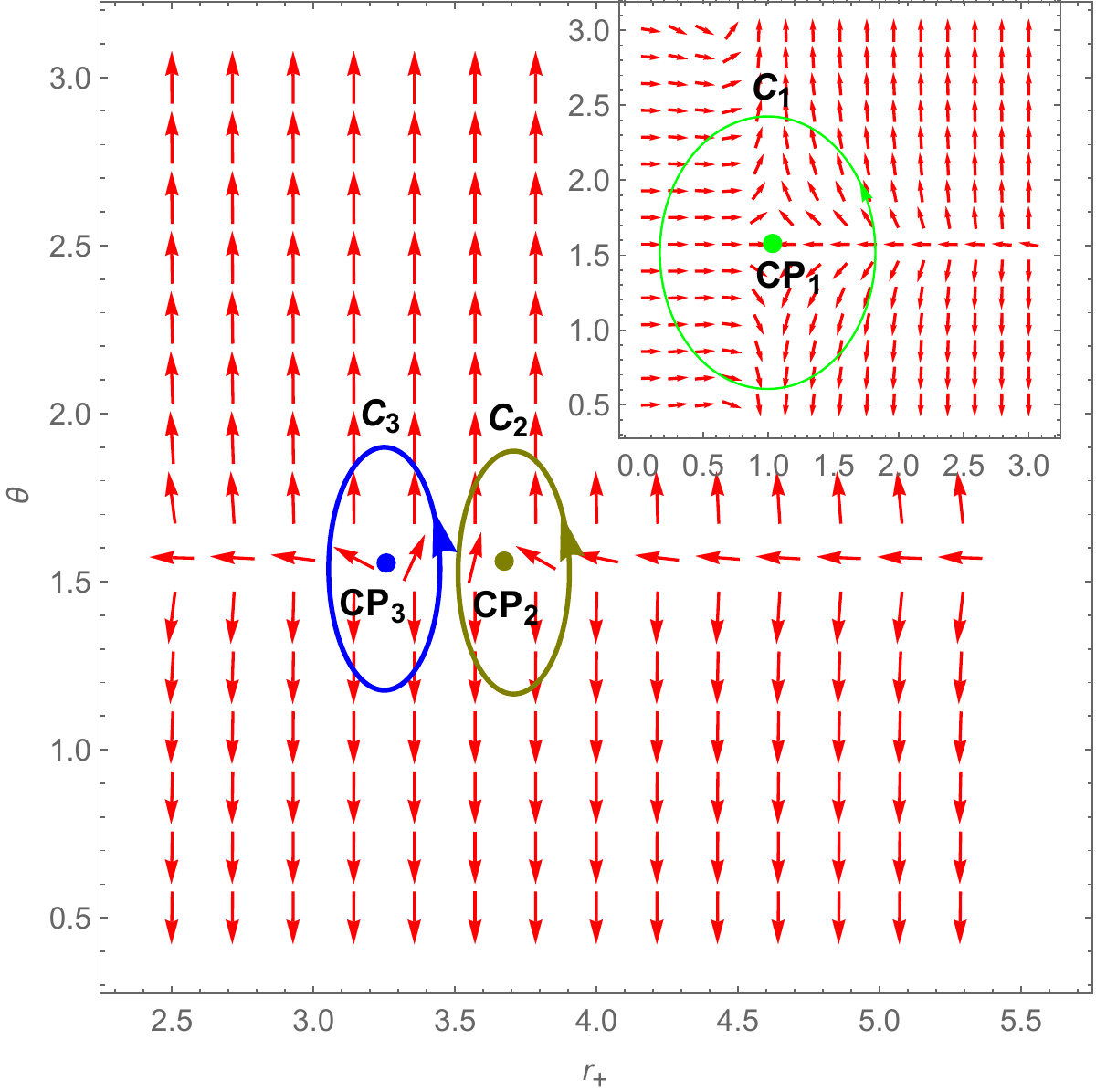} \hspace{-0.5cm}
		\includegraphics[scale = 0.25]{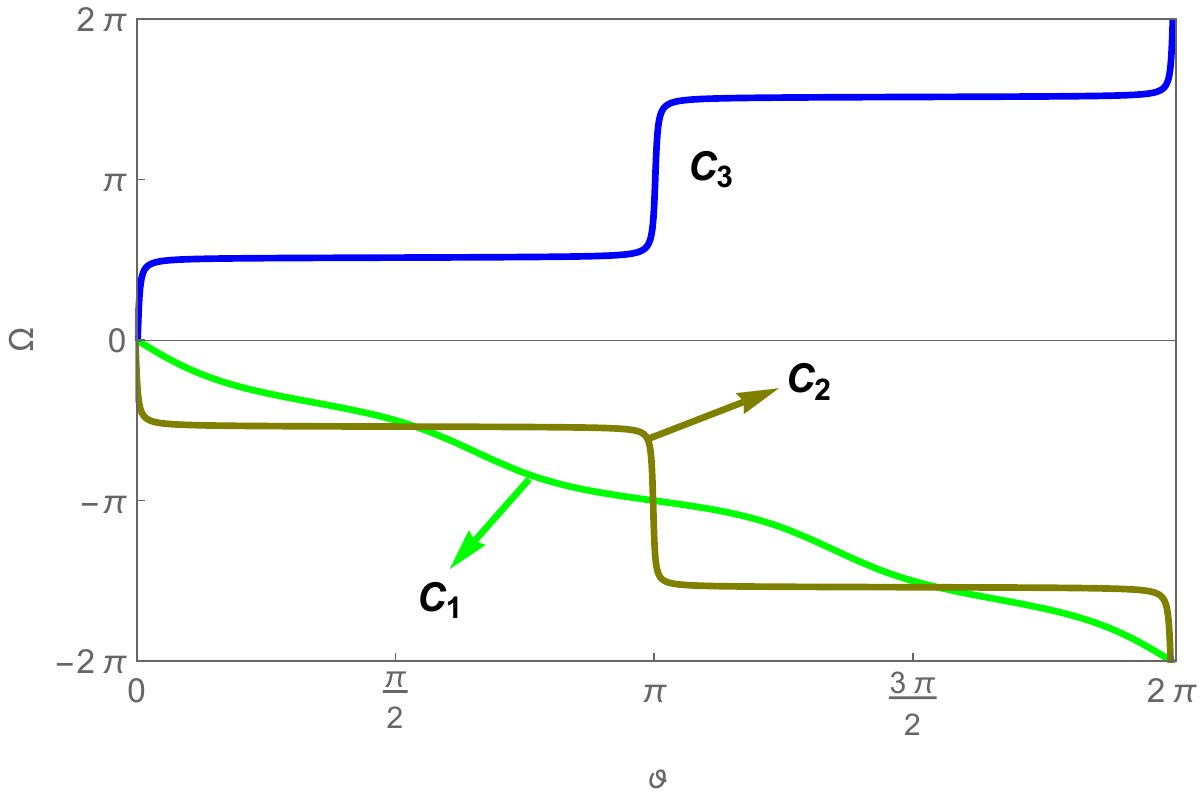} \vspace{-0.5cm}
        \caption{$(d=6, \alpha=1,\beta=0.1,Q_{e}=1,Q_{m}=2)$: \textbf{ Up panel:} The red arrows represent the vector field $n$ for dyonic AdS black holes with QTE in the EGB background. The colored dots are $(r, \theta) =(1.008427, \pi/2 )$, $(r, \theta) =(3.609964, \pi/2 )$ and $(r, \theta) =(3.284875, \pi/2 )$ which represent the critical points $CP_1$, $CP_2$  and $CP_3$, respectively. \textbf{Down panel:} The deflection angle $\Omega (\vartheta)$ as a function of $\vartheta$ for contours $C_1$ (green curve), $C_2$ (brown curve) and $C_3$ (blue curve).}
		\label{fig:WKBL4}
	\end{figure}
\begin{equation}
\begin{aligned}
\Phi & =\frac{1}{4 \pi \sin \theta r_{+}\left(6(d-4)(d-3) \alpha+r_{+}^2\right)} \\
& \times\left[2(3-d)\left(2(d-5)(d-4) \alpha+r_{+}^2\right)\right. \\
& \left.+r_{+}^{8-2 d}\left(Q_m^2+\frac{r_{+}^{4 d} Q_e^2}{\left(r_{+}^{2 d}+8 \beta \Gamma(d-1)^2 r_{+}^4 Q_m^2\right)^2}\right)\right].
\end{aligned}
\end{equation}
The vector field $\phi$ can be resolved into its components $(\phi^{r_{+}}$ and $\phi^{{\theta}})$, through appropriate manipulation, we can derive the outcome represented by
\begin{equation}
\begin{aligned}
 &\phi^{r_{+}}=\frac{\csc \theta}{4 \pi\left(6(d-4)(d-3) \alpha+r_{+}^2\right)^2}[4(d-3) \alpha \\
&\left.+r_{+}^2\right)+2 r_{+}^{8-2 d}\left(\frac{r_{+}^{4 d} Q_e^2}{\left(r_{+}^{2 d}+8 \beta \Gamma(d-1)^2 r_{+}^4 Q_m^2\right)^2}\right. \\
&\left.+Q_m^2\right)-2\left(6(d-4)(d-3) \alpha+r_{+}^2\right) \\
& \times\left(6-2 d-\frac{(d-4) r_{+}^{6+4 d} Q_e^2}{\left(r_{+}^{2 d}+8 \beta \Gamma(d-1)^2 r_{+}^4 Q_m^2\right)^3}\right. \\
&+r_{+}^{6-2 d} Q_m^2\left(+\frac{8 d \beta \Gamma(d-1)^2 r_{+}^{4+4 d} Q_e^2}{\left(r_{+}^{2 d}+8 \beta \Gamma(d-1)^2 r_{+}^4 Q_m^2\right)^3}\right. \\
&+4-d))+\frac{1}{r_{+}^2}\left(6(d-4)(d-3) \alpha+r_{+}^2\right) \\
& \times\left(-2(d-3)\left(2(d-5)(d-4) \alpha+r_{+}^2\right)\right. \\
&+\left.\left.r_{+}^{8-2 d}\left(Q_m^2+\frac{r_{+}^{4 d} Q_e^2}{\left(r_{+}^{2 d}+8 \beta \Gamma(d-1)^2 r_{+}^4 Q_m^2\right)^2}\right)\right)\right].
\end{aligned}
\end{equation}
\begin{equation}
\begin{aligned}
\phi^\theta & =\frac{\cot \theta \csc \theta}{4 \pi r_{+}\left(6(d-4)(d-3) \alpha+r_{+}^2\right)}\left[2(3-d)\right.\\
& \left.\times 2(d-5)(d-4) \alpha+r_{+}^2\right)+r_{+}^{8-2 d} \\
& \left.\left.\times\left(Q_m^2+\frac{r_{+}^{4 d} Q_e^2}{\left(r_{+}^{2 d}+8 \beta \Gamma(d-1)^2 r_{+}^4 Q_m^2\right)^2}\right)\right)\right].
\end{aligned}
\end{equation}
The normalized vector field is defined by
\begin{equation}
n=(\frac{\phi ^r}{\left \| \phi  \right \| },\frac{\phi ^\theta }{\left \| \phi  \right \| } ).
\end{equation}
Next, we investigate the topological charge linked to the critical point. Through analysis and examination of the topological structure, the topological charge $Q$ is assigned a specific value for each critical point, when encircling a single critical point $CP$ with a contour $C$, it is observed that its non-zero topological charge $Q$ signifies significant information associated with the enclosed area. Conversely, if contour $C$ lies outside of critical point, its topological charge becomes zero \cite{cq6,cq7}. If there are many critical points within the contour, the total topological charge $Q$ is obtained by aggregating the individual topological charges associated with each critical point. In the orthogonal plane, we employ the method of Duan's $\varphi$-mapping theory to ensure that the constructed contour exhibits a positive value. We can transform the critical point on the $(S, \theta)$ plane to the $(r, \theta)$  plane using reparameterization. In this case, the critical point  is situated at $(r_0, \frac{\pi}{2})$
\begin{equation}
\left\{\begin{matrix}
r=a\cos\vartheta +r_0,\\ \theta =b\sin\vartheta +\frac{\pi}{2}.
\end{matrix}\right.
\end{equation}
where $\vartheta\subset (0,2\pi)$.
After this transformation \cite{cq4}, we can express the deflection angle
\begin{equation}
\Omega (\vartheta )=\int_{0}^{\vartheta } \epsilon _{ab}n^a \partial_\vartheta n^bd\vartheta.
\end{equation}
Naturally, the topological charge can be determined by the deflection angle as follows:
\begin{equation}
Q=\frac{\Omega (2\pi)}{2\pi}.
\end{equation}
Now, let us investigate the classification of critical points pertaining to AdS black holes. Research has indicated that black holes demonstrate characteristic swallowtail behavior across various dimensions of space-time and undergo small/large BH phase transitions resembling those observed in van der Waals liquid/gas systems \cite{cq3}. Interestingly, in the case of $d = 6$, when the coupling constant $\alpha$ falls within a particular parameter range $(\alpha\geq 0.5)$, a noteworthy phenomenon occurs where the black hole experiences a phase transition characterized by small/intermediate/large BH phases. Additionally, this leads to the emergence of a triple points naturally, signifying that the black hole possesses three critical points. In the case where the coupling constant $\alpha=1$, it can be observed that despite the presence of three critical points, there are indications of the triple point phase transition within the system as suggested by the existence of two swallowtails.\begin{table}[]
	\centering
	\caption{Contours at $Q_{e}=1,Q_{m}=2, \beta=0.1 $ for dyonic AdS black holes with QTE in the EGB background.}
\emph{\begin{tabular}{|c|l|l|l|l|l|l}
  \hline $$ & & $C_1$ & $C_2$ & $C_3$ & $C_4$ \\                                           \hline                                             & $a$    & 0.07    &  0.07      & -           &- \\
($\alpha=0.5, d=5$)                        & $b$    & 0.4     &  0.4       & -           &- \\
                                                   & $r_0$  & 2.658528 &  3.8      &  -          &-\\
\hline                                             & $a$    & 0.15    & 0.15       & -        &-\\
($\alpha=0.1,d=6$)           & $b$    & 0.4     & 0.4        & -         &-\\
                                                   & $r_0$  & 2.57753 & 2.4    & -      &- \\
\hline                                             & $a$    & 0.15    & 0.15       & 0.15        &1.1\\
($\alpha=0.5, d=6$)   & $b$    & 0.4     & 0.4        & 0.4         &0.5\\
                                                   & $r_0$  & 1.489048 & 2.706456 & 1.873569 &2.09775\\
 \hline                                            & $a$    & 0.15    & 0.15       & 0.15        &-\\
($\alpha=1.0, d=6$) &$b$   & 0.4     & 0.4        & 0.4         &-\\
                                                   & $r_0$  & 1.008427 & 3.609964    & 3.284875     &-\\ \hline
\end{tabular}}
\end{table}
We aim to topologically  classify  the critical points for  the dyonic AdS BHs with QTE in the EGB background. By assigning special values to the parameters and performing numerical calculations, we give the critical points of different dimensions in TABLE I. Noted that for $d \neq6$, there is only one critical point for the parameter to be assigned different values, which undergoes a small/large phase transition in the system. For the case of $d = 5$, the characteristics of $\theta$ and radius $r_{+}$ are presented in relation to the vector field $n$ in FIG. 1. Identify a critical point located at $( \theta, r_{+} )$=$(\frac{\pi}{2}, 2.658528 )$. Create two contours, namely $C_{1}$ and $C_{2}$, where $C_1$ should enclose critical point, while no critical points should be enclosed by $C_{2}$ (refer to TABLE II for contour parameters). Similarly, the test found that $d=7,8,9$, and the same situation occurs.\\
Now, let 's analyze the complex situation with triple point $(d = 6)$.\\
$(i)$: $\alpha=0.1, \beta=0.1$\\
The presence of  a conventional critical point for a black hole can be observed at coordinates $(1.711584, \pi/2 )$, with a topological charge of $(Q=Q_{CP_1}=-1)$ in FIG.2, there exists only one critical point for the BH $(\alpha=0.1)$, which undergoes a small/large phase transition in the system.\\
$(ii)$: $\alpha=0.5, \beta=0.1$\\
According to the findings \cite{cq3}, it is explicitly stated that the LBH/ IBH/LBH phase transition occurs occurs at $\alpha = 0.5$  (refer to TABLE.I for other parameter values). The contours $C_1$ , $C_2$, and $C_3$ surround the critical points $CP_1$ $(1.489048, \pi/2)$, $CP_2$ $(2.706456, \pi/2)$, $CP_3$ $(1.873569, \pi/2)$, respectively. The contour $C_4$ encompasses three critical points. From the deflection angle diagram, we can observe that $Q_{CP_1}=Q_{CP_2}=-1$, and $Q_{CP_3}=1$ in FIG.3. Thus, this means that $CP_1$ and $CP_2$ are the conventional critical points, while $CP_3$ is the novel critical point.  We can also find the deflection angle of $C_4$ is equal to -2$\pi$, meaning that the topological charge of the three critical points surrounded by $C_4$ is $-1$, which can also be computed as $Q=Q_{CP_1}+Q_{CP_2}+Q_{CP_3}=-1$. The triple points phenomenon is intriguing due to its manifestation of the distinctive characteristics associated with black hole phase transition. For this occurrence to take place, it is essential for the ${CP_1}$ and ${CP_2}$ to be physical. Unlike the conventional critical point, achieving minimal Gibbs free energy at the novel critical point $CP_3$ is unattainable.\\
$(iii)$: $\alpha=1, \beta=0.1$\\
We find that there are two swallowtails curves in the G-T diagram \cite{cq3} in the range of pressure $CP_3<P<CP_2$,  we will discuss why the system has three critical points and only small/large BH phase transition occurs. The first observation is that the contours $C_1$ ,$C_3$ , and $C_2$ enclose the critical points $CP_1$ $(1.008427, \pi/2)$,$CP_3$ $(3.284875, \pi/2)$,  $CP_2$ $( 3.609964, \pi/2)$ respectively within the vector field in FIG.4. And the contours $C_1$ , $C_2$, and $C_3$ corresponding deflection angle $\Omega (\vartheta)=-2\pi, -2\pi$ and $2\pi$ for $\vartheta=2\pi$, respectively,  this means that the black holes associated topological charge of $ Q_ {CP_1} =Q_ {CP_2} = - $1, and $Q_{CP_3}=1$. As a result, the total topological charge can be determined by $Q=Q_{CP_1}+Q_{CP_2}+Q_{CP_3}=-1$. Naturally, the conventional critical point is denoted as $CP_1$ and $CP_2$, whereas the novel critical point is represented by $CP_3$. As the system undergoes small/large BH phase transition, it can be inferred that only $CP_1$ represents the conventional (physical) critical point. Furthermore, $CP_2$ and $CP_3$ are associated with non-physical critical points. In addition, the findings indicate that the critical point with topological charge of $-1$ can be considered as an indicator of the first order phase transition, whereas the novel critical point does not possess the ability to minimize Gibbs free energy and thus is defined as the critical point associated with non-physical characteristics \cite{ni1}. However, the conventional critical point $CP_2$ here lacks the ability to minimize the Gibbs free energy and is defined as a non-physical critical point. This suggests that the aforementioned conclusion cannot be universally applied to all systems.
\begin{figure}
		\centering
        \includegraphics[scale = 0.26]{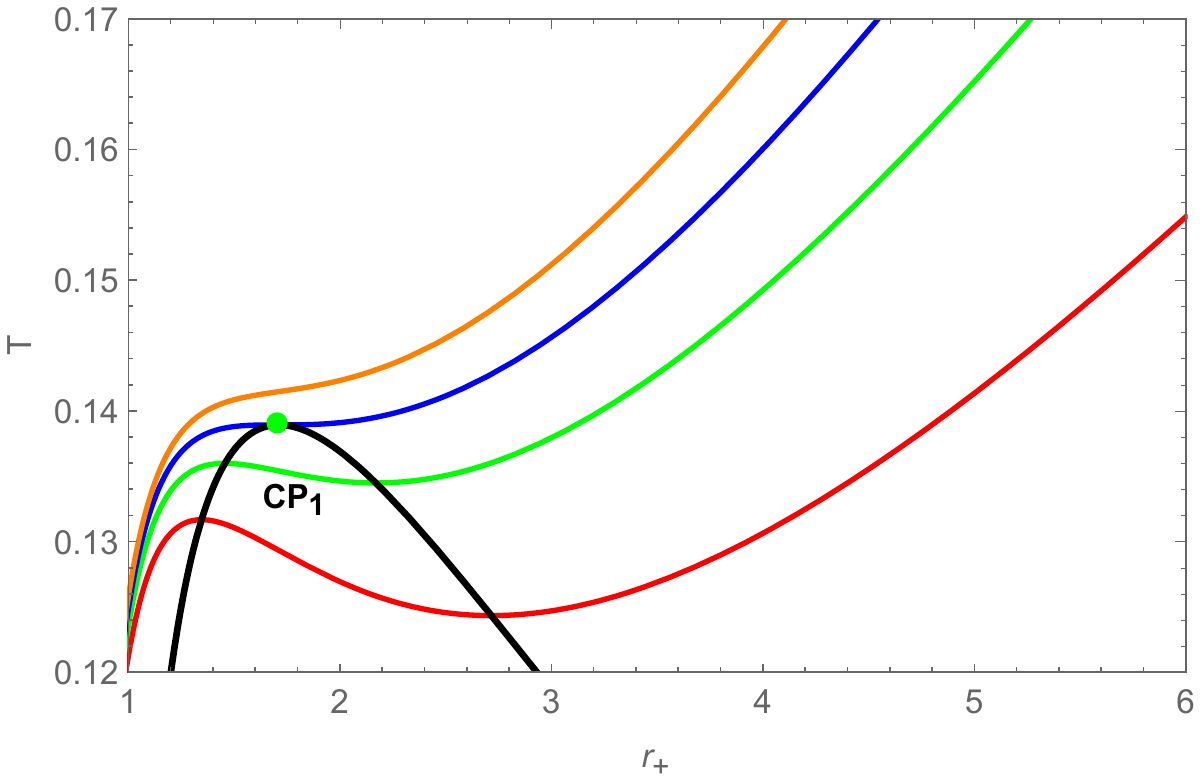} \hspace{-0.5cm}
		\includegraphics[scale = 0.27]{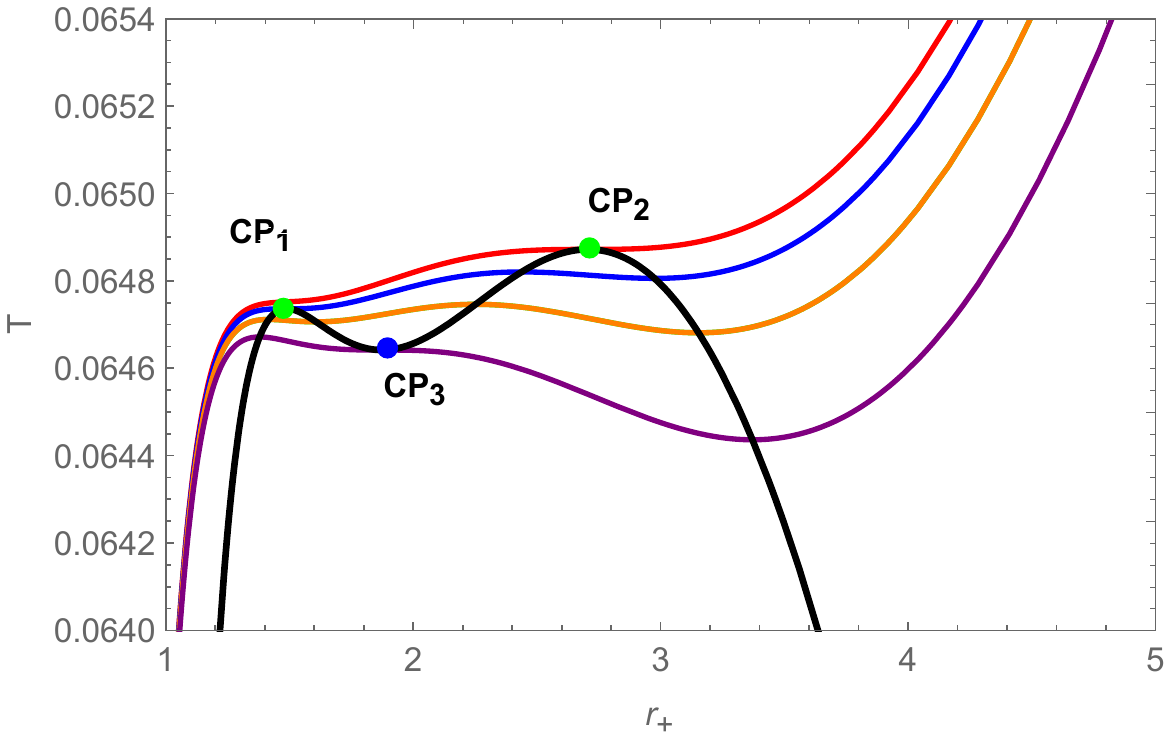} \hspace{-0.5cm}
        \includegraphics[scale = 0.25]{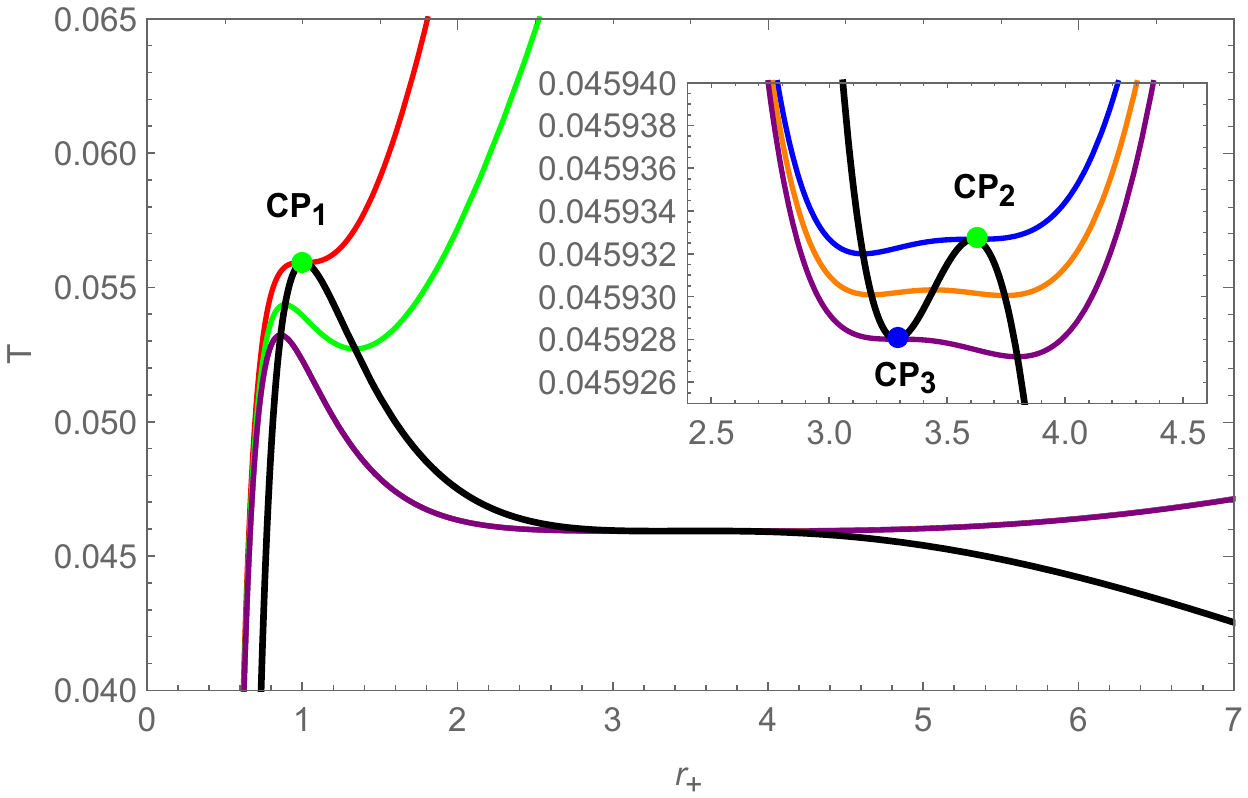} \hspace{-0.5cm}
		
\caption{Isobaric curves (colored solid curves) for dyonic AdS black holes with QTE in the EGB background. shown in the $T-r_{+}$ plane.
The black curve is for the extremal points of the temperature. \textbf{Up panel:}$(\alpha=0.1,colored curve (red:P = 0.02); green:P = 0.025; orange: P = 0.03;  blue: P_{CP1} = 0.027903 )$,\textbf{Middle panel:}$(\alpha=0.5:P_{CP3}=0.06353 (purple) ; P= 0.006469 (orange); P_{CP1}=0.006536 (blue); P_{CP2}=0.0065757 (red))$,\textbf{ Down panel:}$(P_{CP3}=0.03308(purple); P=0.003309(orange); P_{CP2}=0.003311(blue); P=0.025(green); (P_{CP1}=0.050611(red))$}
		\label{fig:WKBL6}
	\end{figure}
\section{Nature of Critical points} \label{sec4}
In this section, we will investigate the nature of the critical points.  We can find some interesting results as follows:\\
$(i)$ $\alpha=0.1, \beta=0.1$\\
\textbf{Up panel:} In FIG.5, we can find each isopiestic curve corresponds to distinct maximum extreme points of temperature, additionally, it is noteworthy that the conventional critical point $(Q_{CP1} = -1)$ coincides with the maximum temperature extreme point. We present a $T-S$ equal-area diagram in FIG.6, here $T_{p}$ denotes the temperature at which the black hole experiences a phase transition. The application of Maxwell's equal area law, we can eliminate the unstable region near $CP_1$ (referred to as the intermediate black hole branch). Moreover, our findings reveal that the isobaric curves are divided into distinct regions with varying phase numbers by the critical temperature, and the arrangement of phases in proximity to the critical point is determined based on these regions. \begin{figure}
		\centering
        \includegraphics[scale = 0.23]{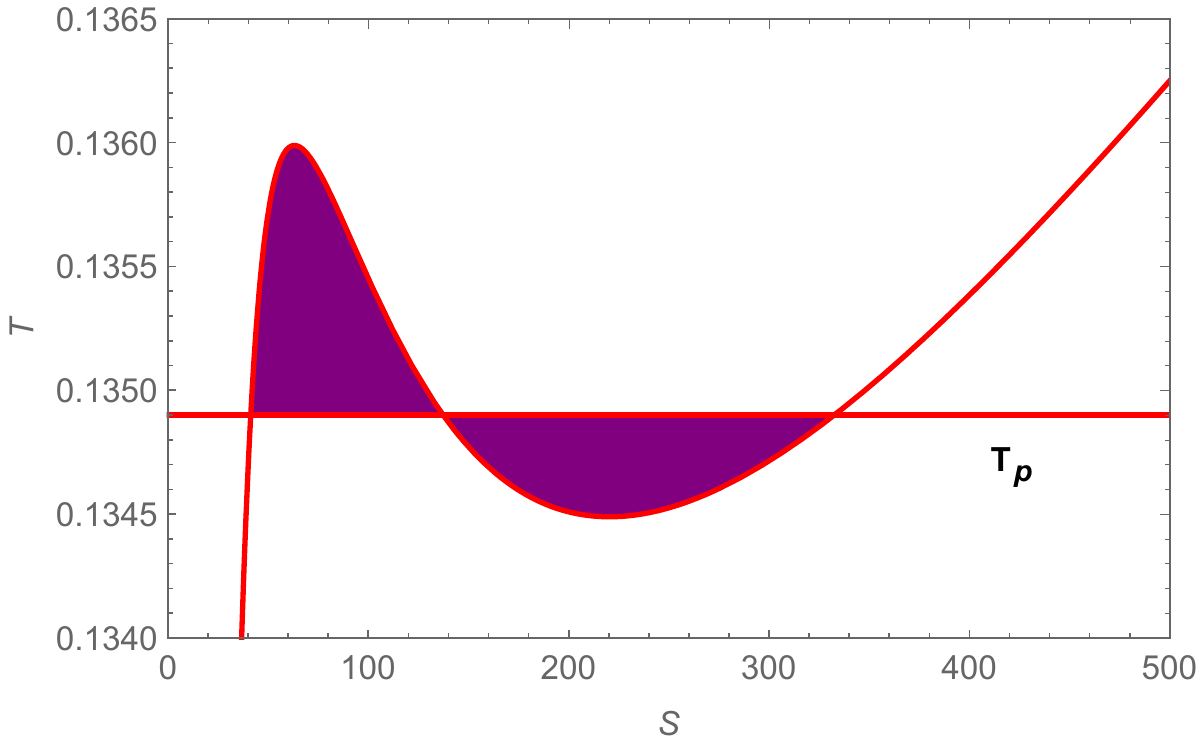} \hspace{-0.5cm}
\caption{$T-S$ diagram of the Maxwell's equal area law $(P=0.025,T_{p}=0.1349, \alpha=0.1)$.}
		\label{fig:WKBL7}
	\end{figure} Interestingly, an increase in pressure $(red\rightarrow orange)$ results in a reduction of isopiestic lines near $CP_1$. Consequently, $CP_1$ can be regarded as a phase annihilation point, noted that $d = 5,7,8,9$ is the same as this case.\\
$(ii)$ $\alpha=0.5, \beta=0.1$\\
\textbf{Middle panel:}
We have observed that the maximum extreme temperature is associated with the green dots, which represent conventional critical points $CP_1$ and $CP_2$. On the other hand, the blue point (novel critical point $CP_3$) accurately represents the minimum extreme temperature.\begin{figure}
		\centering
        \includegraphics[scale = 0.26]{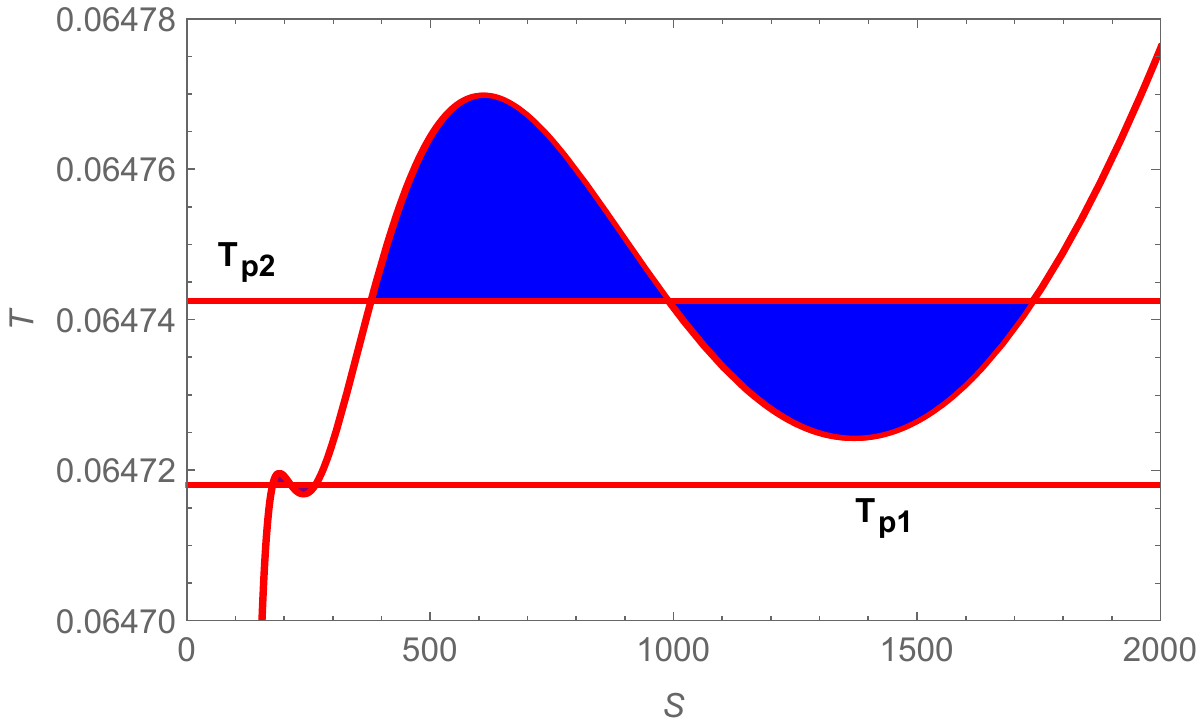} \hspace{-0.5cm}
\caption{$T-S$ diagram of the Maxwell's equal area law $(P=0.006492,T_{p1}=0.064718,T_{p2}=0.0647425, \alpha=0.5)$.}
		\label{fig:WKBL8}
	\end{figure}
\begin{figure}
		\centering
        \includegraphics[scale = 0.26]{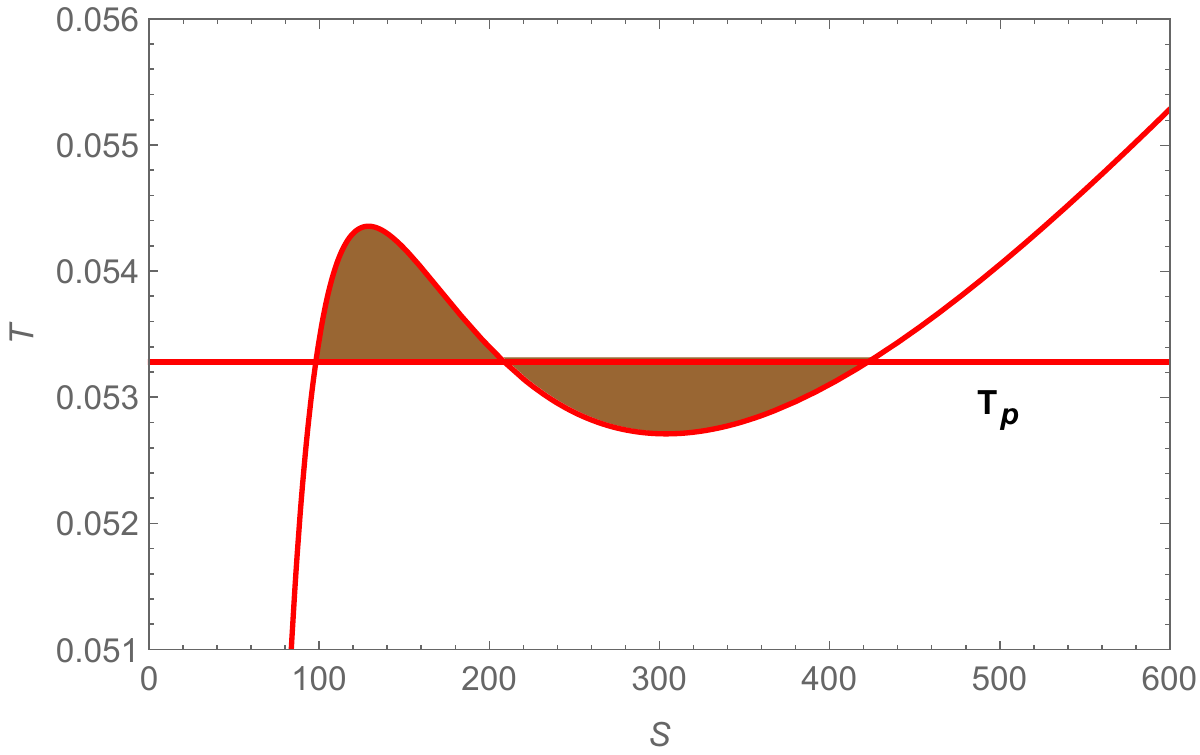} \hspace{-0.5cm}
\caption{$T-S$ diagram of the Maxwell's equal area law $(P=0.025,T_{p}=0.05328, \alpha=1)$.}
		\label{fig:WKBL9}
	\end{figure} We can note that the region of $T-S$ is depicted in FIG.7, with $T_{p1}$ and $T_{p2}$ denoting the temperature of the black hole during small to intermediate and intermediate to large phase transitions, respectively. By applying Maxwell's equal area law, we can effectively eliminate unstable regions near $CP_1$ and $CP_2$; however, this approach may not be successful for unstable regions near $CP_3$. These findings align perfectly with the results \cite{cq4}. Furthermore, as pressure increases (indicated by orange-red), we notice a decrease in the number of regions surrounding $CP_1$ and $CP_2$, as well as a decrease in the number of regions near $CP_3$.  We designate $CP_3$ as the phase generation point while referring to critical points $CP_1$ and $CP_2$ as phase annihilation points.\\
$(iii)$ $\alpha=1, \beta=0.1$\\
Similarly, the green dots ($CP_1$ and $CP_2$) correspond to the point of maximum extreme temperature point, while the blue dot (novel critical point $CP_3$) accurately corresponds to the point of minimum extreme temperature point. Based on research findings \cite{cq3}, it has been observed that there is no occurrence of a first-order phase transition in close proximity to the conventional critical point $CP_2$, which can also be found in \cite{o1}. In FIG.8, we give the temperature $(T_{p}= 0.05328)$ at which the $P = 0.025$ black hole undergoes a phase transition. Notably, the application of Maxwell's law of equal area can effectively remove the unstable region near $CP_2$. However, it is not suitable for indicating a first-order phase transition. Therefore, an alternative classification should be employed, it can be inferred that $CP_2$ may serve as a phase annihilation point.
\section{Dyonic AdS black holes as a topological defect}\label{sec5}
In this section, we intend to examine the topological number and defect curves pertaining to the dyonic black hole featuring QTE within EGB gravity. By incorporating the equations for mass and entropy (Eq. 12 and Eq. 16) into Eq. 31, we can establish the generalized free energy
\begin{equation}
\begin{aligned}
&\mathcal{F} =\frac{1}{32\left(3-4 d+d^2\right) \pi} r_{+}^{-5-d} \mathcal{V}_{d-2} \\
& \quad \times\left[(d-1) Q_m^2 r_{+}^8+2(d-3) r_{+}^{2 d}\right. \\
& \quad \times\left(\left(2-3 d+d^2\right) r_{+}^2\right. \\
& \left.\quad+\left(24-50 d+35 d^2-10 d^3+d^4\right) \alpha+16\pi r_{+}^4 P\right) \\
& \quad+(d-1) Q_e^2 r_{+}^8{ }_2 F_1 \\
& \left.\quad \times\left(1, \frac{d-3}{2(d-2)}, \frac{7-3 d}{4-24},-8 \beta Q_m^2 r_{+}^{4-2 d} \Gamma(d-1)^2\right)\right] \\
& -\frac{r_{+}^{d-2} \pi^{\left(\frac{d-1}{2}\right)}\left(1+2(d-3)(d-2) \alpha r_{+}^{-2}\right)}{2 \tau \Gamma\left(\frac{d-1}{2}\right)},
\end{aligned}
\end{equation}
the components of the vector $\phi$ is given by
\begin{equation}
\begin{aligned}
\phi^{r_{+}}&=  \frac{1}{16 \Gamma\left(\frac{d-1}{2}\right)} \pi^{\frac{(d-3)}{2}} r_{+}^{-6-d}\left(-r_{+}^8 Q_m^2+r_{+}^{2 d}\right. \\
& \times\left(32 P \pi r_{+}^4+2(-3+d)(-2+d)\right. \\
& \times\left(r_{+}^2+(d-5)(d-4) \alpha\right) \\
& -\frac{8(d-2) \pi r_{+}\left(r_{+}^2+2(d-4)(d-3) \alpha\right)}{\tau} \\
& \left.\left.-\frac{r_{+}^8 Q_e^2}{r_{+}^{2 d}+8 r_{+}^4 \beta \Gamma(d-2)^2 Q_m^2}\right)\right),
\end{aligned}
\end{equation}
\begin{equation}
\phi^{\Theta}=-\cot \Theta \csc \Theta .
\end{equation}
 The expression for $\tau$ can be obtained by setting $\phi^r+$ equal to zero
\begin{equation}
\tau=\frac{8(d-2) \pi r_{+}^{2 d+1}\left(r_{+}^2+2 \tilde{\alpha}\right) \chi_2}{-r_{+}^8 Q_m^2 \chi_2+r_{+}^{2 d}\left(-r_{+}^8 Q_e^2+2\left(\chi_3 r_{+}^2+(-5+d) \chi_1\right) \chi_2\right)},
\end{equation}
here
\begin{equation}
\chi_2=r_{+}^{2 d}+8 r_{+}^4 \beta \Gamma(d-2)^2 Q_m^2,
\end{equation}
and
\begin{equation}
\chi_3=6+(-5+d) d+16 P \pi r_{+}^2.
\end{equation}
Next, we will detect the topological number of black holes in different dimensions.
\subsection{The small/large phase transition $(d=5,7,8,9)$.}\label{sec5}
Since black holes $(d=5,7,8,9)$ only have  small / large  phase transition, accompanied by a critical point, we only study the case of $d = 5$. In this case, the pressure is below its critical threshold, and similar to the phase transitions observed in van der Waals liquid/gas systems, small/large phase transitions occur within the black hole. To ascertain the zero points, we consider the unit vectors by setting $\Theta=\pi / 2$ in $n^1$ and equating it to zero. By considering $\tau / r_0=15$, the unit vector $n$ is depicted on the $\left(r_+ / r_0, \Theta\right)$ plane in FIG.9, we find the presence of a zero point at $r_+ / r_0=34.0601$. The vector changes its direction once it passes through a zero point along $\Theta=\pi / 2$.By examining the zero point $ZP_{1}$, we observe that the winding number of $ZP_{1}$, the sum of these winding numbers yields a topological number $W=+1$.If we consider $\tau / r_0=20$ in FIG.10, we find that three zero points are located in $r_+ / r_0=1.30526$,2.5324, 34.0601.We observe that the winding number of $ZP_{2}$ and $ZP_{4}$ is positive $(+1)$ while that of $ZP_{3}$ is negative $(-1)$. The sum of these winding numbers yields a topological number $W=+1-1+1=+1$. In FIG. 11, we present a single zero point vector diagram when $\tau / r_0=38$. Similarly, the same results appear when $d = 7,8,9.$, the system shares a similar topological classification as the charged RN-AdS black holes  \cite{we1}. The relationship between $r_{+}$ and $\tau$ is presented in FIG.12, as previously derived in Eq.36, revealing the existence of two turning points.
In  \cite{we1}, the authors established an annihilation point and a generation point by considering the condition $\frac{d^2 \tau}{d r_+^2}>0$ and $\frac{d^2 \tau}{d r_+^2}<0$ at a specific point $\tau_c$, respectively. Based on this, we can observe the presence of a point of generation and a point of annihilation.
\begin{figure}
		\centering
        \includegraphics[scale = 0.3]{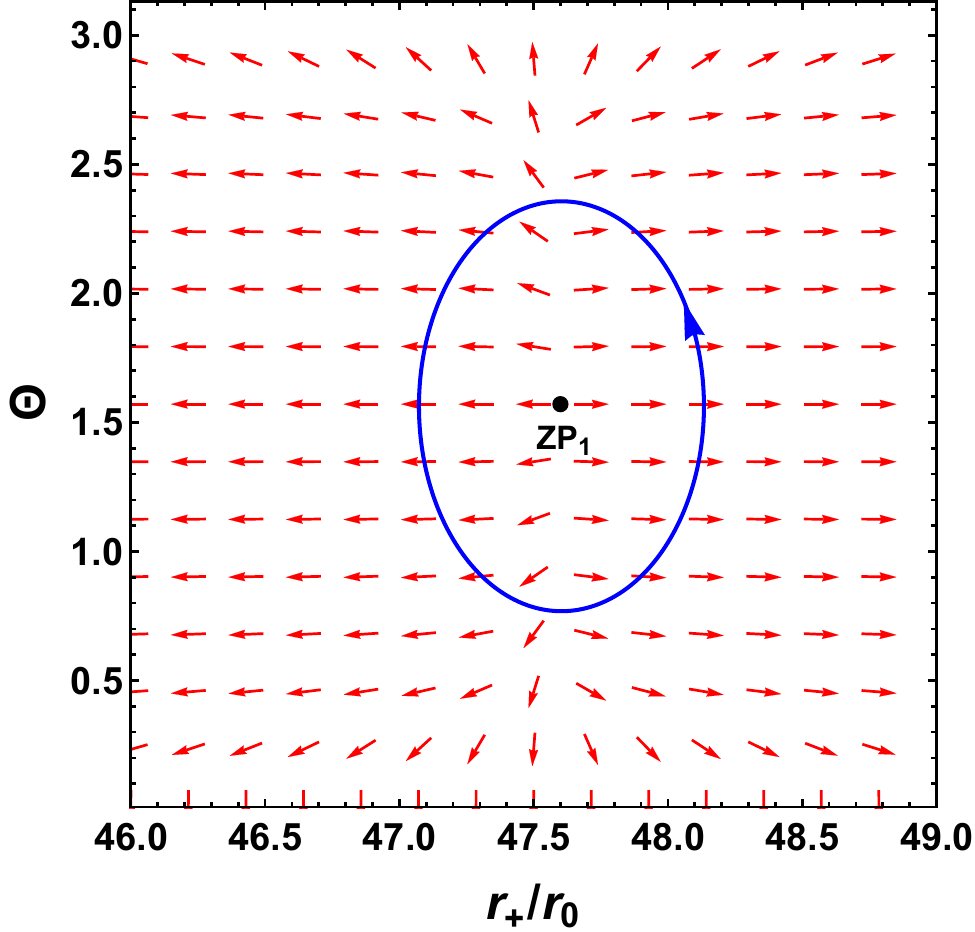} \hspace{-0.5cm}
\caption{Unit vector $n=\left(n^1, n^2\right)$ shown in $\Theta$ vs $r_{+} / r_0$ plane for $\operatorname{Pr}_0^2=0.001$. The black dot represents zero point ($(\tau / r_0=15,\alpha/r_0=0.5,\beta/r_0=0.1)$.}
		\label{fi1}
	\end{figure}
\begin{figure}
		\centering
		\includegraphics[scale = 0.3]{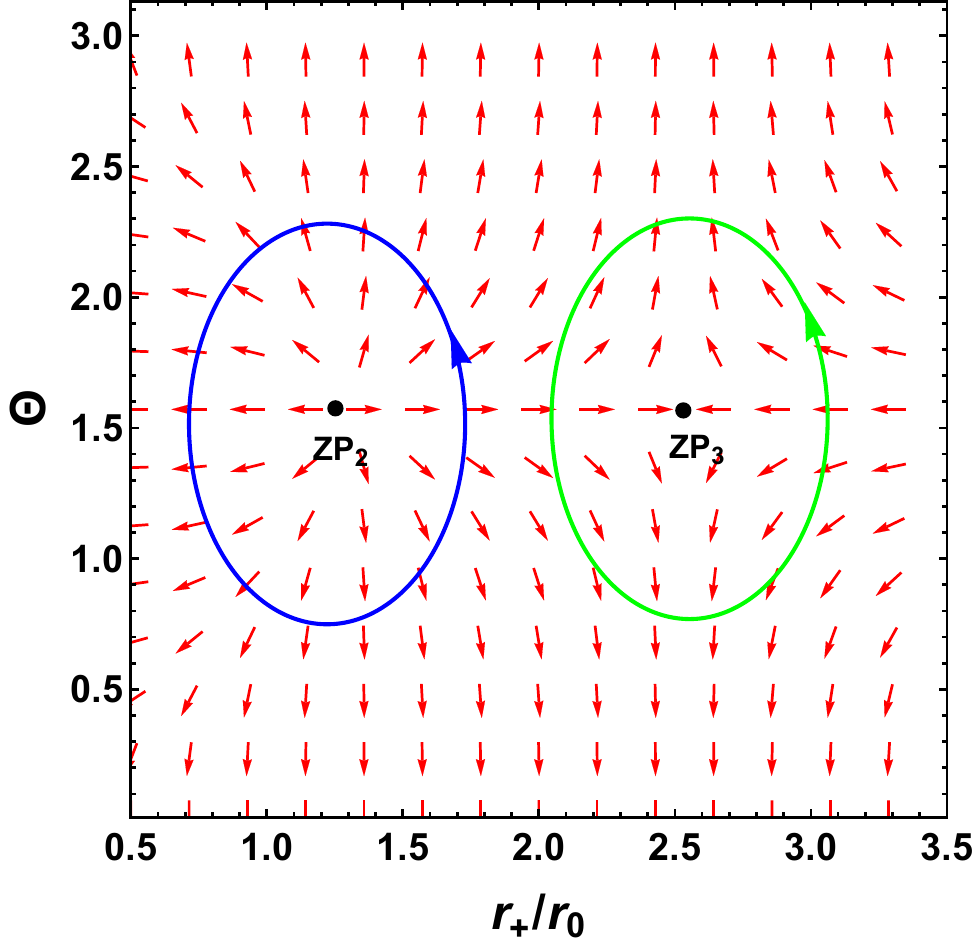} \hspace{-0.5cm}
		\includegraphics[scale = 0.3]{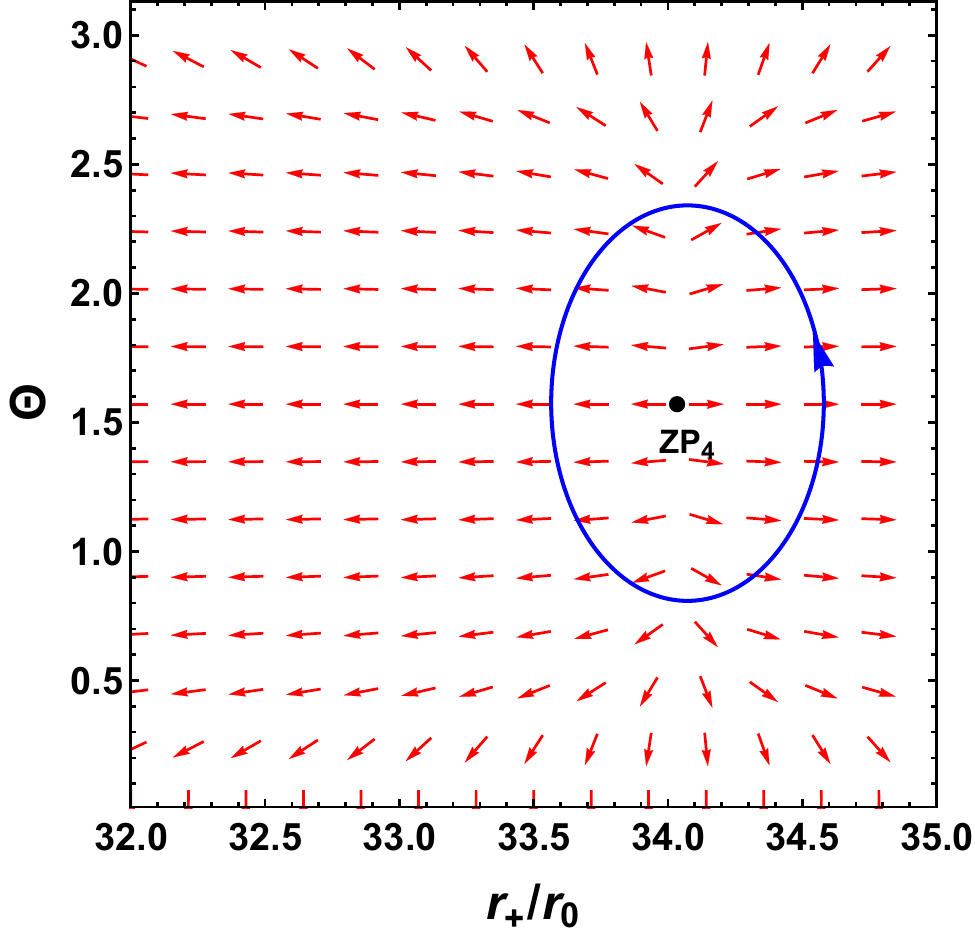} \vspace{-0.5cm}
        \caption{Unit vector $n=\left(n^1, n^2\right)$ shown in $\Theta$ vs $r_{+} / r_0$ plane for $\operatorname{Pr}_0^2=0.001$. The black dot represents zero point ($(\tau / r_0=20,\alpha/r_0=0.5,\beta/r_0=0.1)$.}
		\label{fig:WKBL4}
	\end{figure}
\begin{figure}
		\centering
        \includegraphics[scale = 0.3]{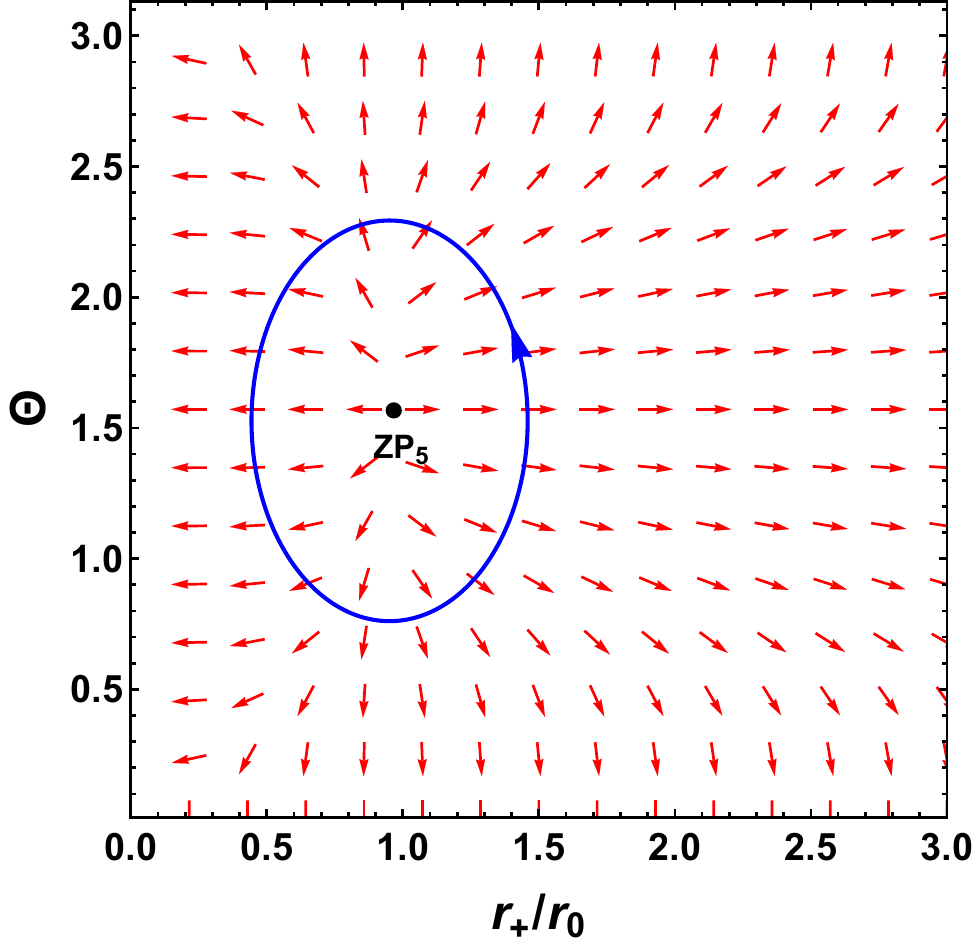} \hspace{-0.5cm}
\caption{Unit vector $n=\left(n^1, n^2\right)$ shown in $\Theta$ vs $r_{+} / r_0$ plane for $\operatorname{Pr}_0^2=0.01$. The black dot represents zero point ($(\tau / r_0=38,\alpha/r_0=0.5,\beta/r_0=0.1)$.}
		\label{fi1}
	\end{figure}
\begin{figure}
		\centering
        \includegraphics[scale = 0.4]{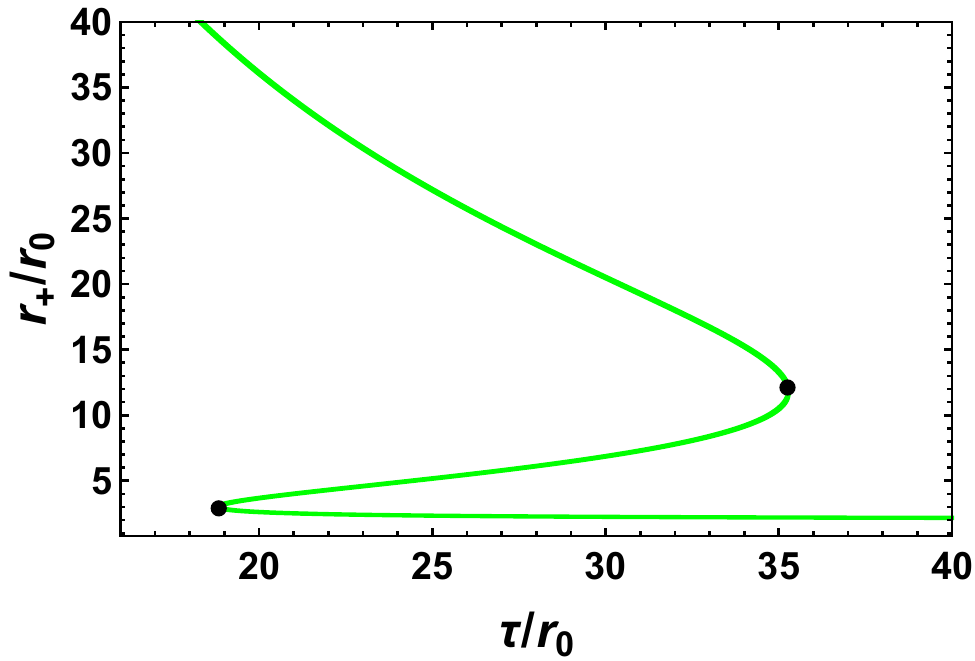} \hspace{-0.5cm}
\caption{Defect curves in the $r_h / r_0-\tau / r_0$ plane with $P r_0^2=0.001,\alpha/r_0=0.5,\beta/r_0=0.1$.}
		\label{fi1}
	\end{figure}

\subsection{The triple points $(d=6)$.}\label{sec5}
Now, we will involve examining the topological number  pertaining to the dyonic black hole featuring $\alpha/r_0=0.5$ and $\beta/r_0=0.1$. For the sake of simplicity, we consider three cases where the pressure $Pr^2_0$ is equal to 0.006, 0.0064, and 0.05. Here, $r_0$ represents an arbitrary length scale determined by a cavity that encompasses the black hole.\\
$(i)$ $Pr^2_0=0.006$\\
We find that this result is consistent with the results of other dimensional black holes $(d=5,7,8,9)$ in Figs.13-14. Under the background of $\Theta=\pi / 2$, there are three zero  points located at $r_+ / r_0=1.08325,2.54691$, and $4.87928$, the sum of these winding numbers yields a topological number $W=+1-1+1=+1$. In addition, we can observe the presence of a point of generation and a point of annihilation.\\
$(ii)$ $Pr^2_0=0.0064$\\
\begin{figure}
		\centering
        \includegraphics[scale = 0.3]{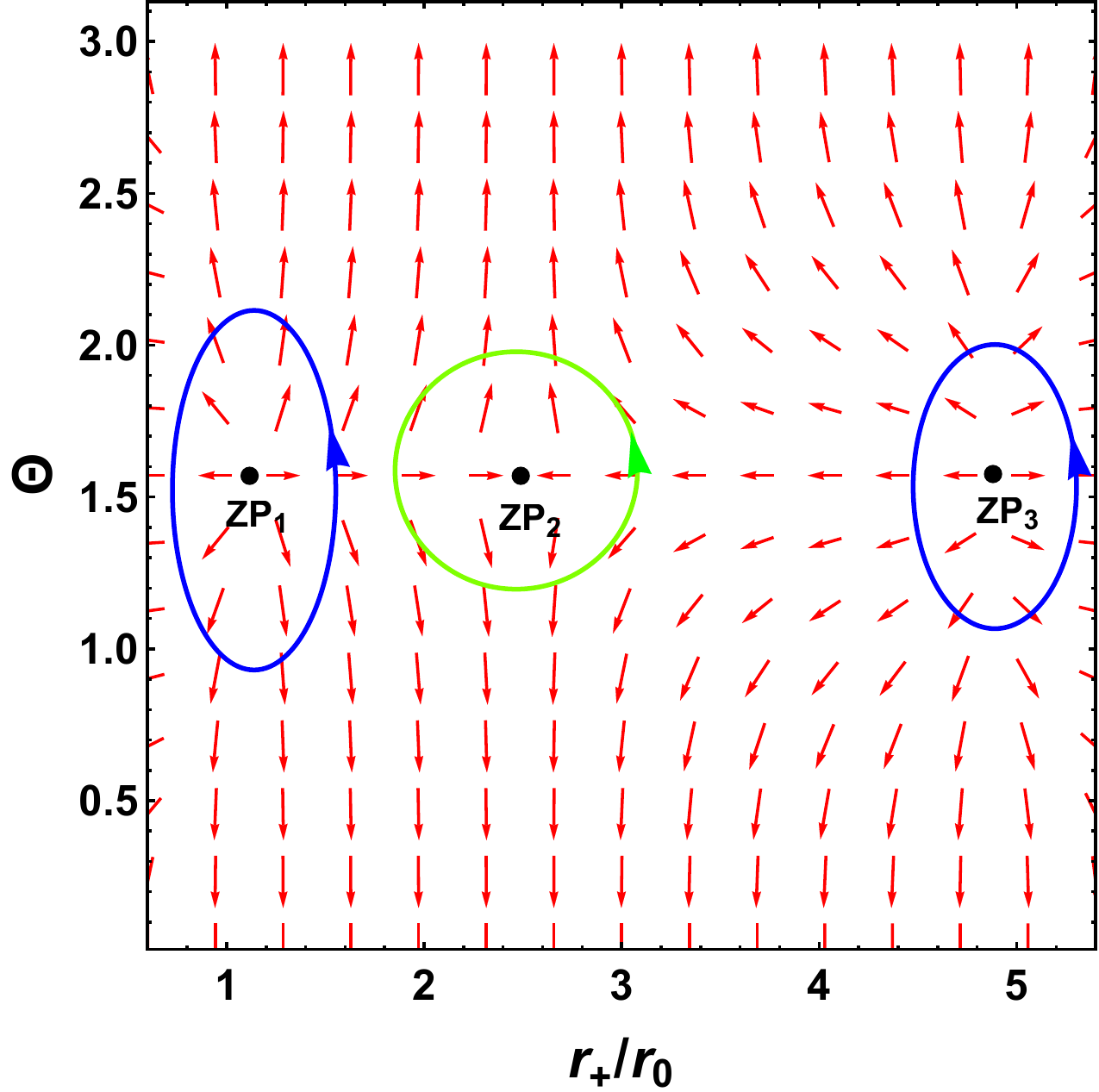} \hspace{-0.5cm}
\caption{Unit vector $n=\left(n^1, n^2\right)$ shown in $\Theta$ vs $r_{+} / r_0$ plane for $\operatorname{Pr}_0^2=0.006$. The black dot represents zero point ($(\tau / r_0=15.6)$.}
		\label{fi1}
	\end{figure}

\begin{figure}
		\centering
        \includegraphics[scale = 0.4]{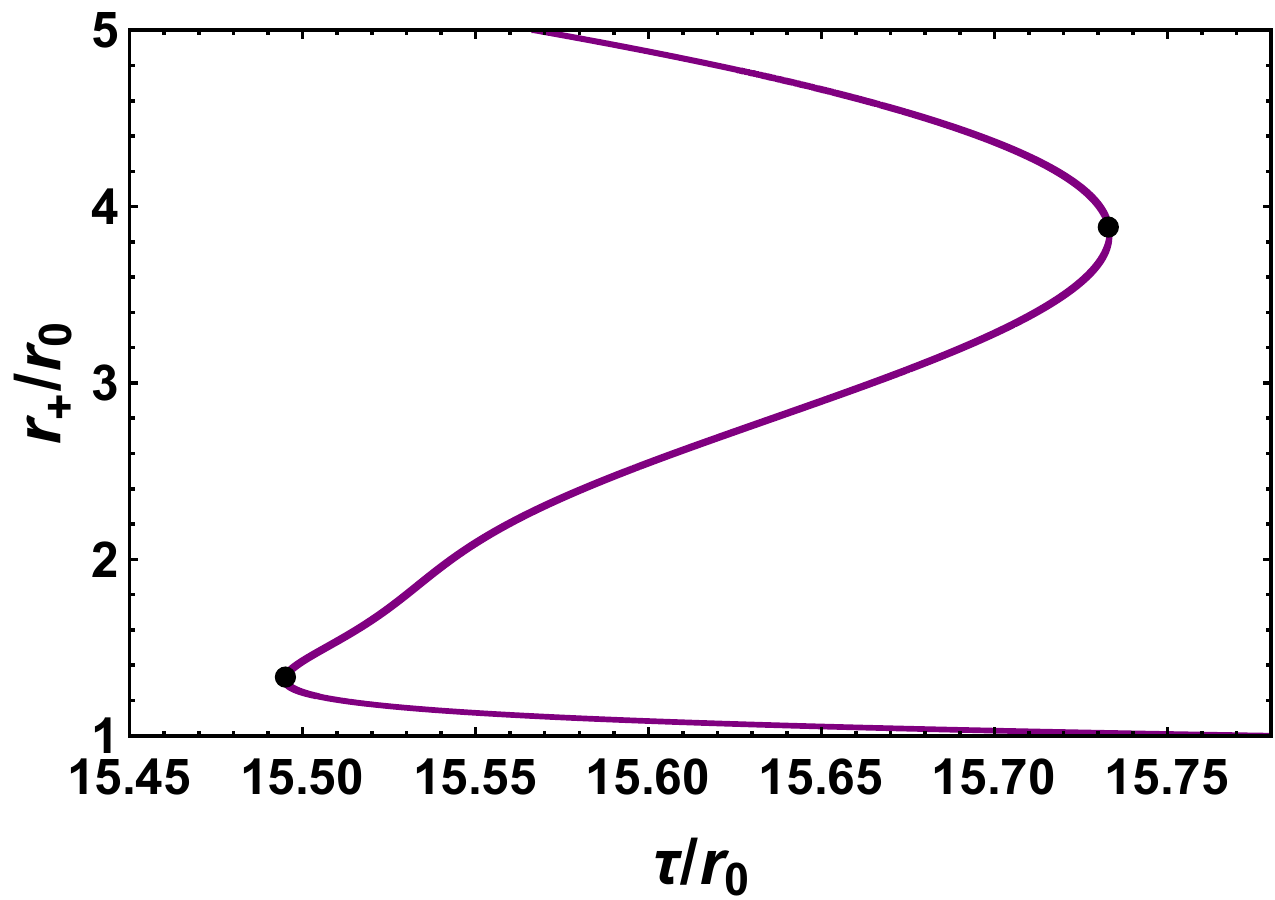} \hspace{-0.5cm}
\caption{Defect curves in the $r_h / r_0-\tau / r_0$ plane with $P r_0^2=0.006$.}
		\label{fi1}
	\end{figure}
\begin{figure}
		\centering
        \includegraphics[scale = 0.3]{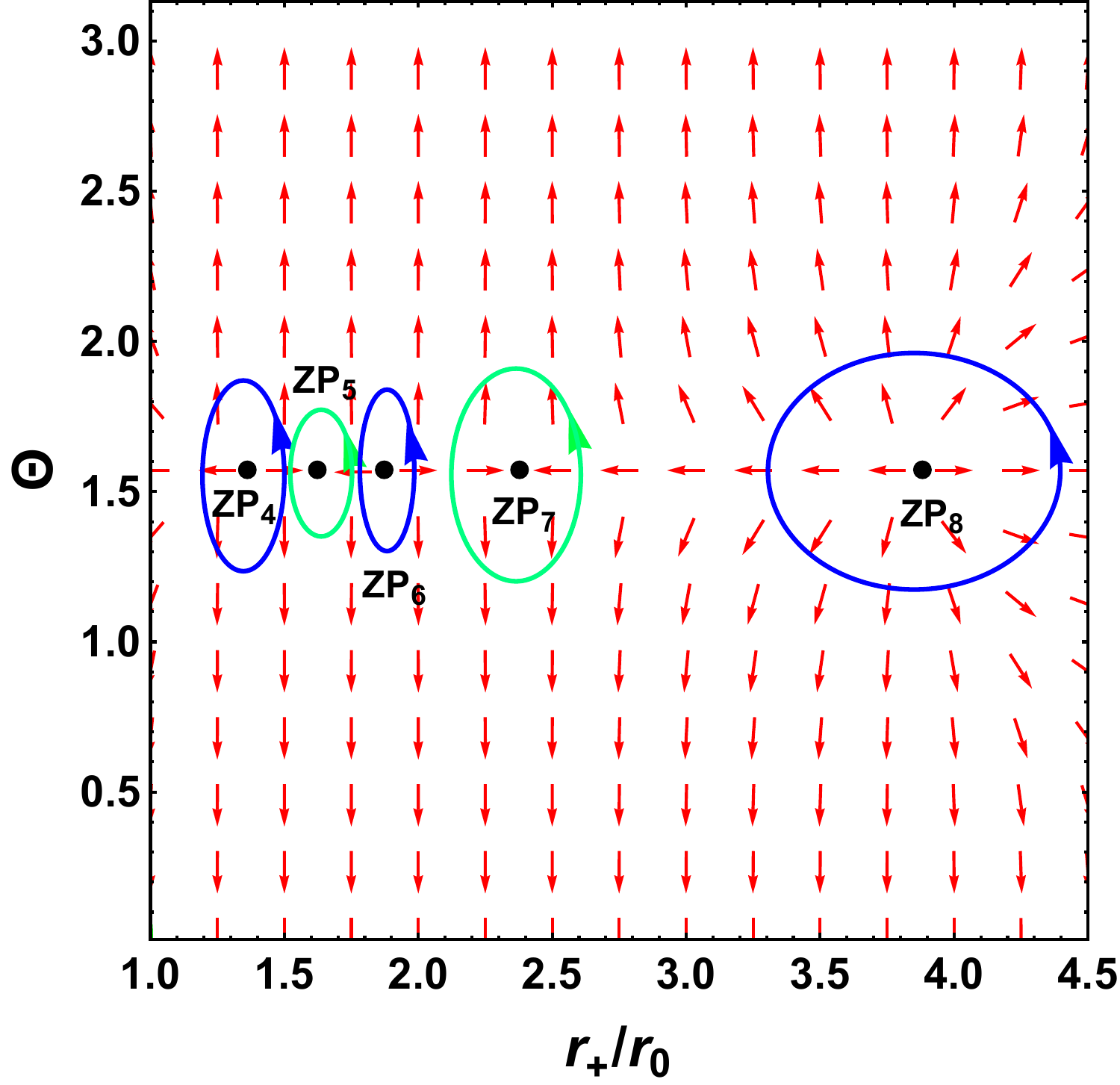} \hspace{-0.5cm}
\caption{Unit vector $n=\left(n^1, n^2\right)$ shown in $\Theta$ vs $r_{+} / r_0$ plane for $\operatorname{Pr}_0^2=0.0064$. The black dot represents zero point ($(\tau / r_0=15.468)$.}
		\label{fi1}
	\end{figure}Here, we adjust the pressure to achieve a value of $P r_0^2=0.0064$, the additional features related to defects and defect curves  will be observed.
We observe that the $\left(r_+ / r_0, \Theta\right)$ plane in FIG.15 exhibits five zero points at $r_+ / r_0=1.29839,1.51839, 1.88713, 2.40912$, and $3.83608$, as indicated by the unit vector $n$. Similarly, we can observe that the zero points $ZP_{4}$, $ZP_{6}$, and $ZP_{8}$ possess a winding number of $+1$, whereas $ZP_{5}$ and $ZP_{7}$ exhibit a winding number of $-1$. By summing up all the winding numbers,  the topological number to be $W=+1-1+1-1+1=+1$. \begin{figure}
		\centering
        \includegraphics[scale = 0.4]{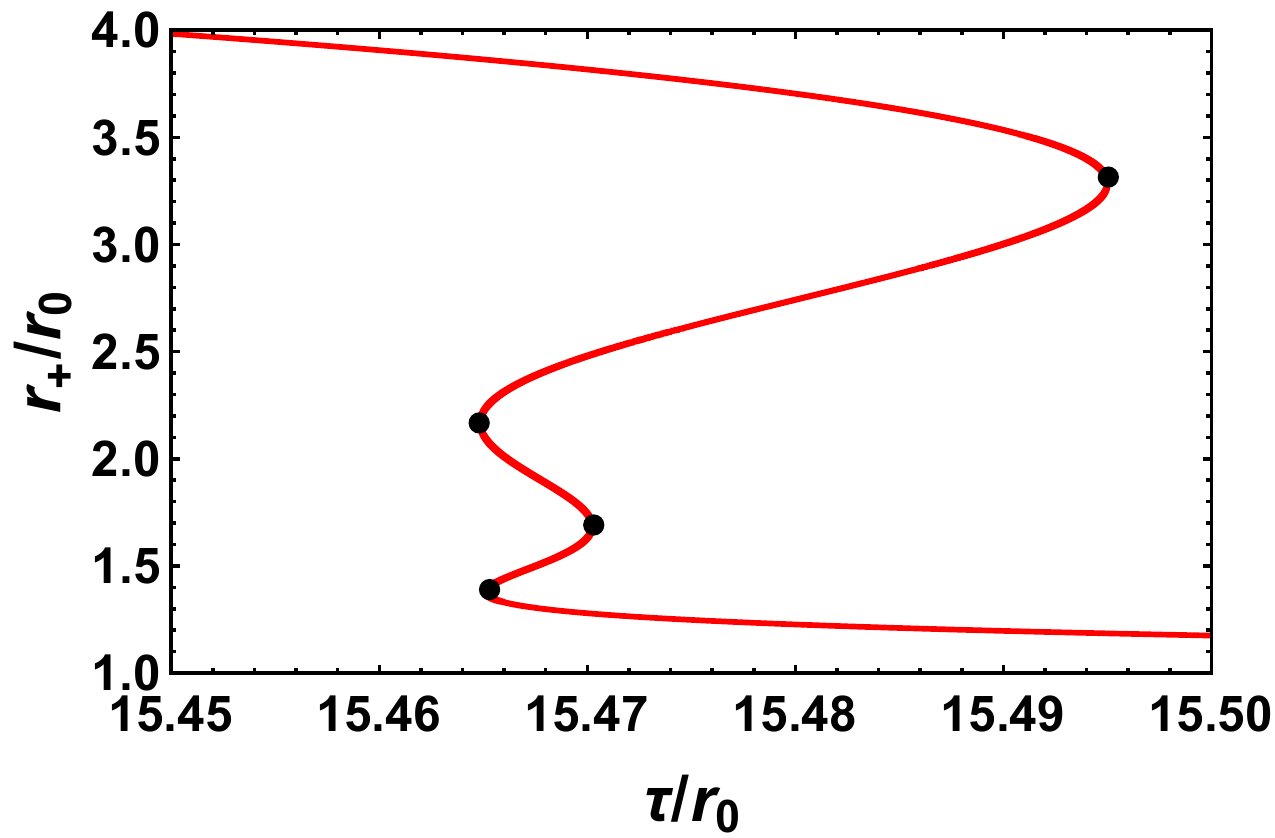} \hspace{-0.5cm}
\caption{Defect curves in the $r_h / r_0-\tau / r_0$ plane with $P r_0^2=0.0064$.}
		\label{fi1}
	\end{figure} Interestingly, it can be observed that the system exhibits two points of generation and two points of annihilation in FIG.16, this implies that with an increase in pressure, there will be alterations in both the generation point and the annihilation point.  The total topological number, however, remains unchanged at $Pr_0^2=0.006$. For dyonic black holes with QTE, the $r_h / r_0-\tau / r_0$ plane manifests two separate curves, \cite{ni2}. However, our findings indicate that the continuity of the curve in the $r_h / r_0-\tau / r_0$ plane is evident, this implies that the characteristics of $r_h / r_0-\tau / r_0$ are influenced by the coupling constants $\alpha$ and $\beta$.
\\
$(iii)$ $Pr^2_0=0.05$\\
For this case,  the pressure exceeds its critical threshold, resulting in the absence of any phase transition. We depict the unit vector in FIG.17 when $\tau / r_0=2$. It is evident that the zero point exists at coordinates $\left(\frac{r_h}{r_0}, \Theta\right) = (2.01201, \frac{\pi}{2})$.  The winding number of the zero point $Z P_{9}$ is +1. According to FIG.18, it can be observed that the system does not produce any generation points or annihilation points.
\begin{figure}
		\centering
        \includegraphics[scale = 0.3]{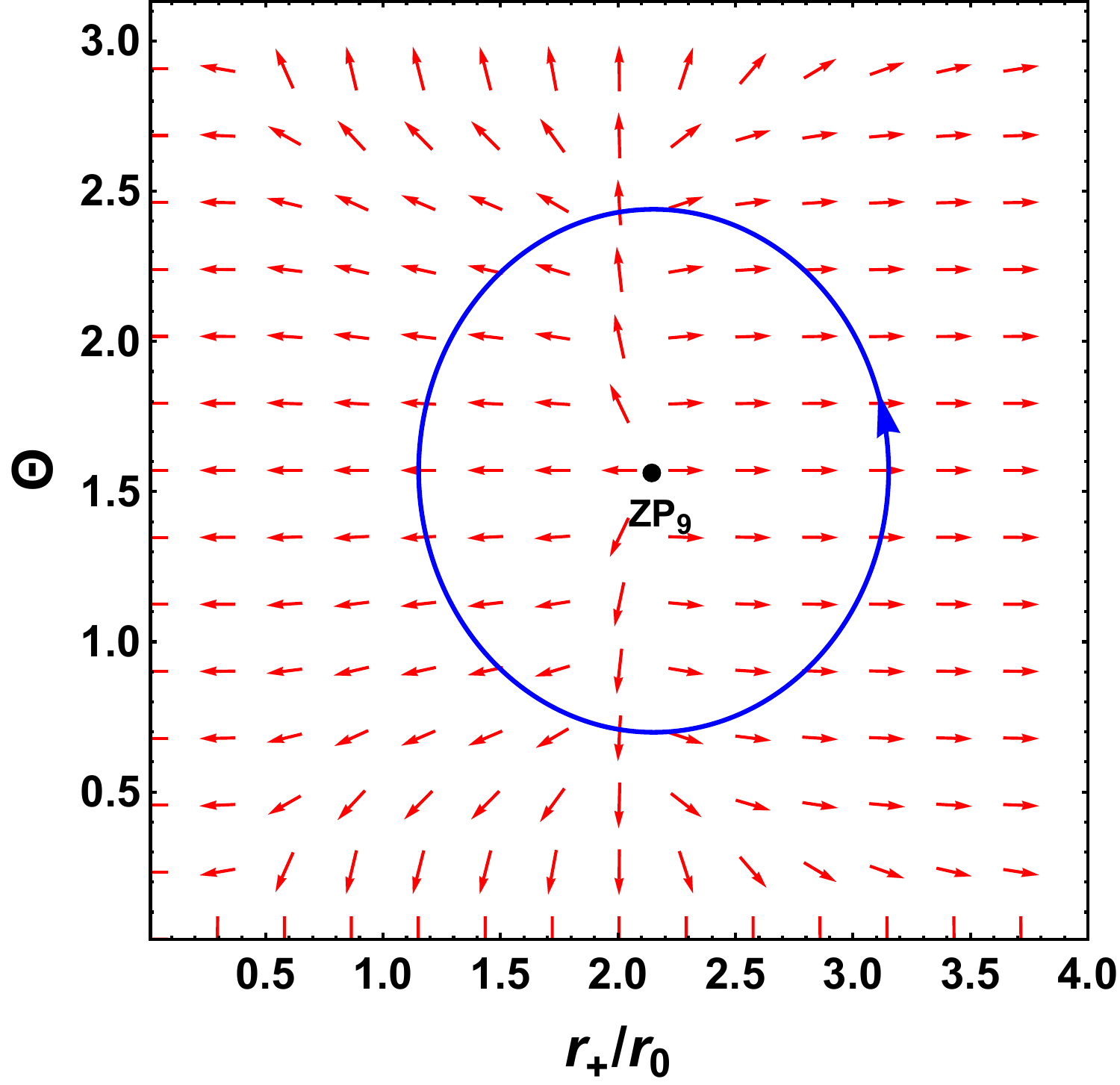} \hspace{-0.5cm}
\caption{Unit vector $n=\left(n^1, n^2\right)$ shown in $\Theta$ vs $r_{+} / r_0$ plane for $\operatorname{Pr}_0^2=0.05$. The black dot represents zero point ($(\tau / r_0=10)$.}
		\label{fi1}
	\end{figure}
\begin{figure}
		\centering
        \includegraphics[scale = 0.3]{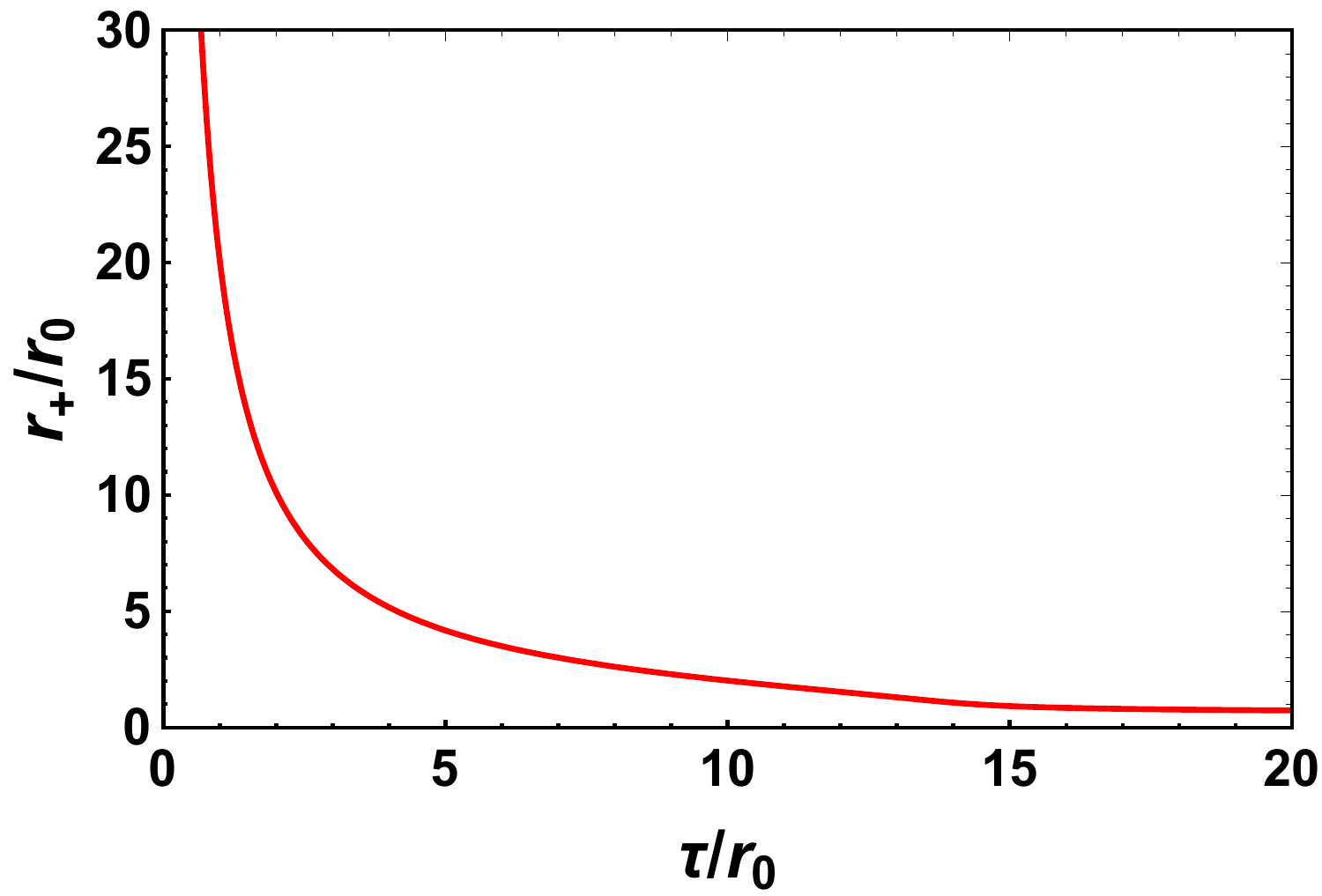} \hspace{-0.5cm}
\caption{Defect curves in the $r_h / r_0-\tau / r_0$ plane with $P r_0^2=0.05$.}
		\label{fi1}
	\end{figure}

Of specific significance, the defects observed on this curve exhibit a winding number of $+1$, resulting in a topological number $W = +1$. Therefore, the dyonic AdS black holes with QTE in EGB gravity belong to the same topological class as the charged RN-AdS black holes \cite{we1}.
\section{Conclusions}\label{sec6}
In summary, when considering a particular value for the GB coupling constant $\alpha$, the phase structure of 6-dimensional dyonic AdS black holes becomes more intricate, with the appearance of triple points \cite{cq3}.  We initially investigate the topological charge associated with the critical point, followed by an analysis of dyonic AdS black holes as a thermodynamic space's topological defect, in order to assess the topological  structure of dyonic AdS black holes. The primary highlights can be summarized as follows:\\
	
$(i)$ We  find that when the coupling constant satisfies $\alpha = 0.5$ and $1$, the system will appear  triple points, but the phase structure is completely different. More specifically, the three critical points can be classified into two conventional (physical) critical points, $( CP_{1}$ and $CP_{2})$, as well as a novel critical point $( CP_{3} )$, the novel critical point is not a physical point because it does not minimize the Gibbs free energy. When a = 1, the system has only one traditional ( physical ) critical point $(CP_{1})$. In both cases, the total topological charge is $-1$. \\
	
$(ii)$ From the isobaric curve and critical temperature in FIG.4, we can observe that the conventional critical point is located at the maximum extreme point of temperature, and the unstable region in close proximity to it can be eliminated through the application of Maxwell's equal area law. It is found that the critical point $CP_{3}$ is no longer applicable. Noted that the novel critical point appears at the minimum extreme point of temperature. The results also reveal a decrease in the number of phases along the isobaric curve at the conventional critical point as pressure increases $(\alpha=1)$, while the number of phases increases at the novel critical point. Therefore, we define the $CP_{2}$  as the phase annihilation point $(\alpha=1)$.\\
$(iii)$ We regard dyonic AdS black holes as a topological defect in the thermodynamic space, wherein  the black hole has a generation point and annihilation point upon the pressure being below critical threshold. The black hole exhibits two points of generation and annihilation as the pressure increases. When the pressure surpasses a critical threshold, there are no production or annihilation points. However, the total topological number of black holes in different dimensions is $1$, the system shares a similar topological classification as the charged RN-AdS black holes \cite{we1}.
\vspace{5mm}
	\begin{acknowledgments}
We would like to thank the anonymous referees for their valuable comments on improving our paper. This work is supported by  the Doctoral Foundation of Zunyi Normal University of China (BS [2022] 07, QJJ-[2024]-203), and the National Natural Science Foundation of China (Grant Nos. 12265007). Also, H. H. is grateful to Excellence project FoS UHK 2203/2025-2026 for the financial support. B.C.L is grateful to Excellence project FoS UHK 2212 / 2021-2022 for the financial support.
	\end{acknowledgments}

\end{document}